\documentclass[
    twocolumn,
    usenames,
    svgnames,
    dvipsnames,
    x11names,
    hyperlink,
]{aastex631}

\usepackage{microtype}
\usepackage{comment}

\let\tablenum\relax
\usepackage{siunitx}
\usepackage{amsmath}
\usepackage[capitalise,noabbrev]{cleveref}

\DeclareSIUnit\pc{pc}
\DeclareSIUnit\kpc{kpc}

\usepackage{lipsum}

\newcommand{\rmd}{\mathrm{d}}
\newcommand{\rme}{\mathrm{e}}

\shortauthors{Gossan et al.}
\begin{document}

\title{Optimizing the third generation of gravitational-wave observatories for Galactic astrophysics}

\correspondingauthor{Sarah E. Gossan}
\email{sarah.gossan@gmail.com}

\author[0000-0002-8138-9198]{Sarah E. Gossan}
\affiliation{Canadian Institute for Theoretical Astrophysics, University of Toronto, 60 St. George Street, Toronto, ON M5S 3H8, Canada}

\author[0000-0001-9018-666X]{Evan D. Hall}
\affiliation{Department of Physics and LIGO Laboratory, Massachusetts Institute of Technology, Cambridge, MA 02139, USA}

\author[0000-0001-6573-7773]{Samaya M. Nissanke}
\affiliation{GRAPPA, Anton Pannekoek Institute for Astronomy and Institute of High-Energy Physics, University of Amsterdam, Science Park 904, 1098 XH Amsterdam, The Netherlands}
\affiliation{Nikhef, Science Park 105, 1098 XG Amsterdam, The Netherlands}

\begin{abstract}
Gravitational-wave (GW) astrophysics is a rapidly expanding field, with plans to enhance the global
ground-based observatory network through the addition of larger, more sensitive observatories:
the Einstein Telescope and Cosmic Explorer. These observatories will allow us to peer deeper into the
sky, collecting GW events from farther away and earlier in the universe. Within our own Galaxy,
there is a plethora of interesting GW sources, including core-collapse supernovae, phenomena in
isolated neutron stars and pulsars, and potentially novel sources. As GW observatories are
directionally sensitive, their placement on the globe will affect the observation of Galactic
sources. We analyze the performance of one-, two-, and three-observatory networks, both for sources
at the Galactic center, as well as for a source population distributed over the Galactic disk. We find
that, for a single Cosmic Explorer or Einstein Telescope observatory, placement at near-equatorial
latitudes provides the most reliable observation of the Galactic center. When a source population
distributed over the Galactic disk is considered, the observatory location is less impactful,
although equatorial observatories still confer an advantage over observatories at more extreme
latitudes. For two- and three-node networks, the longitudes of the observatories additionally become
important for consistent observation of the Galaxy.
\end{abstract}


\section{Motivation}\label{sec:motivation}
The era of gravitational-wave (GW) astronomy has well and truly begun, with more than 67 compact
object binaries observed since Advanced LIGO (Laser Interferometric Gravitational-Wave
Observatory) saw ``first light'' in September
2015~\citep{LIGOScientific:2018mvr,LIGOScientific:2020ibl}. While the coalescence of compact objects
has long been the most promising source for ground-based GW observatories due to their
(comparatively high) expected rates, there exists a plethora of prospective sources in the
tens-of-hertz to kilohertz regime right in our very own cosmic backyard -- the Milky Way galaxy.
These diverse sources range from core-collapse supernovae (see,
e.g.,~\citet{Gossan:2015xda,Abdikamalov:2020jzn,LIGOScientific:2019ryq}, and references therein),
accreting neutron stars in binaries~\citep{Holgado:2017vut}, fallback accretion onto neutron stars
to form black holes~\citep{Piro:2012ax,Melatos:2014pka}, spin-down of isolated neutron
stars~\citep{Andersson:2010ufc,Alford:2012yn}, Thorne--\.{Z}ytkow Objects
(T\.{Z}Os;~\citet{DeMarchi:2021vwr}), accretion-induced collapse of white
dwarfs~\citep{Abdikamalov:2009aq}, pulsar glitches~\citep{vanEysden:2008pd}, and magnetar
flares~\citep{Andersson:2010ufc}. These sources are complementary to the thousands of Galactic
detached and accreting white dwarf binaries, and neutron star and black hole binaries observable in
the millihertz regime by the space-based LISA mission (see,
e.g.,~\citet{Korol:2020hay,Korol:2021pun} and references therein). In the first three science runs
of the LIGO, Virgo, and KAGRA observatories, there have been a number of targeted searches for GWs from
the Galactic Center (see, e.g.,~\citet{LIGOScientific:2013wcb,Piccinni:2019zub}) as well as for continuous waves from young supernova remnants
(e.g.~\citet{Millhouse:2020jlt,LIGOScientific:2021mwx}), accreting millisecond
pulsars~\citep{LIGOScientific:2021ozr,LIGOScientific:2020gml}, known
pulsars~(e.g.~\citet{LIGOScientific:2017hal}), and X-ray binaries~(e.g.,~\citet{Zhang:2020rph}), to
name but a few. 

To date, no GW emission from a Galactic source has been detected, namely because of their
comparatively weaker emission to that from inspiralling and merging compact binaries. However, the
significant increase of sensitivity in the next generation of laser-interferometeric detectors offers
an unprecedented opportunity to discover, observe, and measure GW emission from the entire zoo of
Galactic GW sources. This paper focuses whether the prospects for detecting Galactic source
populations can be improved through the combination of different placements and orientations of the
proposed ground-based third-generation GW observatories.

Previous studies (see, e.g., \citet{Nissanke:2011ax,Raffai:2013yt,Hu:2014bpa,Szolgyen:2017hfh,Mills:2017urp,Zhao:2017cbb,Hall:2019xmm,Ackley:2020atn,Borhanian:2020ypi}) have considered how
the placement and orientation of future generations of ground-based GW observatory networks might be
optimized with respect to a number of figures of merit.  Explicitly, these were: (1) the network's
ability to reconstruct the source polarization, (2) the extent to which the network can localize
the sky position of the source, and (3) the accuracy and precision with which the network can
reconstruct the parameters for a standard compact binary source. Of the Galactic GW sources
detectable by ground-based GW observatories mentioned above, the majority are expected to be accompanied by 
electromagnetic and/or neutrino counterparts (see, e.g., \citet{Andersson:2010ufc,Piro:2012ax,Piro:2013voa,Piro:2012cv,Piro:2014kfa,Haskell:2015iia,Haskell:2015jra,Warren:2019lgb,DeMarchi:2021vwr,LIGOScientific:2021ozr}), from which one can localize and have temporal
information about the event if the source is transient in nature. For such sources, sky localization
from GWs alone is not critical, and we can perform a targeted GW search for such sources with
increased sensitivity. 

In this study, we work to establish the optimum placement and orientation of the network of current
and third-generation ground-based GW observatories in the context of observing Galactic sources. We do
so by introducing the network antenna power, a figure of merit useful for establishing the
detectability of Galactic GW source populations, many of which are unmodeled. We consider different
combinations of one-, two-, and three-observatory networks with different possible orientations and
locations across the globe. The major ground-based third-generation GW observatory effort in Europe
is the Einstein Telescope~\citep{Punturo:2010zz,Maggiore:2019uih,ETDesignUpdate2020}, while
in the United States it is the Cosmic Explorer~\citep{Evans:2021gyd}. Many science goals
involving GWs are best realized if these observatories are operated simultaneously as part of a
global network, which would be enhanced by the addition of a third-generation observatory in
Australia~\citep{Bailes:2019oma} or potentially elsewhere in the Southern Hemisphere. We note that
most of the power from Galactic sources is at very low frequencies (below \SI{100}{\Hz}) and beyond
the reach of current generation ground-based observatories, and hence the placement of new
third-generation GW observatories is critical. 

The rest of this paper is organized as follows. In Section~\ref{sec:basicidea}, we outline the
observatory networks considered, and introduce the figures of merit used in this study. In
Section~\ref{sec:results-onedet}, we present results on the performance of a single observatory for 
Galactic sources, before investigating the same for two- and three-observatory networks in
Section~\ref{sec:results-detnetwork}. We then discuss the implications of these results as well as
the limitations of our methods in Section~\ref{sec:discussion}, before presenting our conclusions in
Section~\ref{sec:conc}.

\section{Basic idea}

\subsection{Observatory Antenna Patterns}\label{sec:basicidea}

Today's large-scale GW observatories use laser-interferometric detectors to sense
passing gravitational waves \citep{Adhikari:2013kya}.
Interferometric detectors are not uniformly sensitive across the celestial sky. The antenna power receivable by a particular detector (given GW emission from
a source located in particular region of the sky) is dependent on its location, orientation,  and
detector geometry. 

The placement of detector $d$ may be characterized relative to the Earth-fixed frame using four
angles: latitude $\Lambda$ and longitude $\lambda$ of the detector's vertex, opening angle $\zeta$
between the detector's two arms, and orientation angle $\gamma$, measured counterclockwise from due
East to the bisector of the detector's two arms.  Similarly, the wave frame (for a given source) can
be defined relative to the same Earth-fixed frame with three angles at any given time $t$: the right
ascension $\alpha$, the declination $\delta$, and the polarization angle $\psi$.

Following \citet{Jaranowski:1998qm}, \citet{Arnaud:2001my}, \citet{Schutz:2011tw}, and references therein, we compute
the antenna sensitivities $F_{+,\times;\,d}\,(\alpha,\delta;t)$ at any time $t$ using
\begin{align}
    F_{+;\,d}\,(\alpha,\delta; t) &= \sin\zeta\, [ a_{d}(\alpha,\,\delta;t) \cos 2\psi \notag\\
    &\hphantom{=} \hspace{5em} + b_{d}(\alpha,\delta;t) \sin 2\psi  ]\,,\notag\\
    F_{\times;\,d}(\alpha,\delta; t) &= \sin\zeta\, [ b_{d}(\alpha,\delta; t) \cos 2\psi \notag\\
    &\hphantom{=} \hspace{5em} - a_{d}(\alpha,\delta;t) \sin 2\psi  ]\,,
\end{align}
where
\begin{align}
    a_{d}(\alpha,\delta;t) &= \frac{1}{16}\sin 2\gamma (3 - \cos 2\Lambda) (3 - \cos 2\delta) \cos 2 \mathfrak{h}(\alpha,\lambda;t)  \notag\\
    &- \frac{1}{4} \cos 2\gamma  \, \sin{\Lambda} \, (3 - \cos 2\delta ) \, \sin 2 \mathfrak{h}(\alpha,\lambda;t)  \notag\\
    &+ \frac{1}{4} \sin{2\gamma} \, \sin 2\Lambda \, \sin 2\delta  \, \cos{\mathfrak{h}(\alpha,\lambda;t)}  \notag\\
    &- \frac{1}{2} \cos 2\gamma \, \cos \Lambda \, \sin 2\delta  \, \sin \mathfrak{h}(\alpha,\lambda;t) \notag\\
    &+ \frac{3}{4} \sin 2\gamma \, \cos^{2} \Lambda \, \cos \delta \,
\end{align}
and
\begin{align}
    b_{d}(\alpha,\delta;t) &= \cos 2\gamma \, \sin \Lambda \, \sin \delta \cos 2 \mathfrak{h}(\alpha,\lambda;t) \notag\\
    &+ \frac{1}{4}\sin 2\gamma \, (3 - \cos 2\Lambda) \sin \delta \, \sin 2\mathfrak{h}(\alpha,\lambda;t) \notag\\
    &+ \cos 2\gamma \, \cos\Lambda \, \cos\delta \, \cos \mathfrak{h}(\alpha,\lambda;t) \notag\\
    &+\frac{1}{2}\sin 2\gamma \, \sin 2\Lambda \, \cos \delta \, \sin \mathfrak{h}(\alpha,\lambda;t)\,.
\end{align}
Here, $\mathfrak{h}(\alpha,\lambda;t)$ is the local hour angle, and 
$d$ denotes the dependence of these expressions on the particulars of the 
detector. 

Averaging then over the polarization angle, one may obtain a `source-independent' measure of 
the antenna sensitivity \citep{Arnaud:2001my}:
\begin{equation}
    \overline{F}_{d}(\alpha,\delta;t) = \sin{\zeta} \; \sqrt{\frac{a_{d}^{2}(\alpha,\delta;t) + b_{d}^{2}(\alpha,\delta;t)}{2}}\,.
    \label{eq:Fbar}
\end{equation}
For an observatory such as Cosmic Explorer which comprises a single \textsf{L}-shaped detector, its
antenna pattern $\overline{F}_{\mathsf{L}}(\alpha,\delta;t)$ is given directly by \cref{eq:Fbar}
with $\zeta = \SI{90}{\degree}$. For a $\triangle$-shaped observatory such as the Einstein
Telescope, comprising three detectors each with $\zeta = \SI{60}{\degree}$ and with orientations
satisfying $\gamma_1 = \gamma_2 - \SI{120}{\degree} = \gamma_3 - \SI{240}{\degree}$, the combined
antenna pattern for all detectors is
\begin{equation}
    \overline{F}_{\triangle}(\alpha,\delta;t) = \sqrt{\overline{F}_{1}^{\,2} + \overline{F}_{2}^{\,2} + \overline{F}_{3}^{\,2}}\,,
    \label{eq:Fbar_triangle}
\end{equation}
where indices $i = 1,2,3$ denote detector number.

\begin{figure}[tph]
    \centering
    \includegraphics[width=\columnwidth]{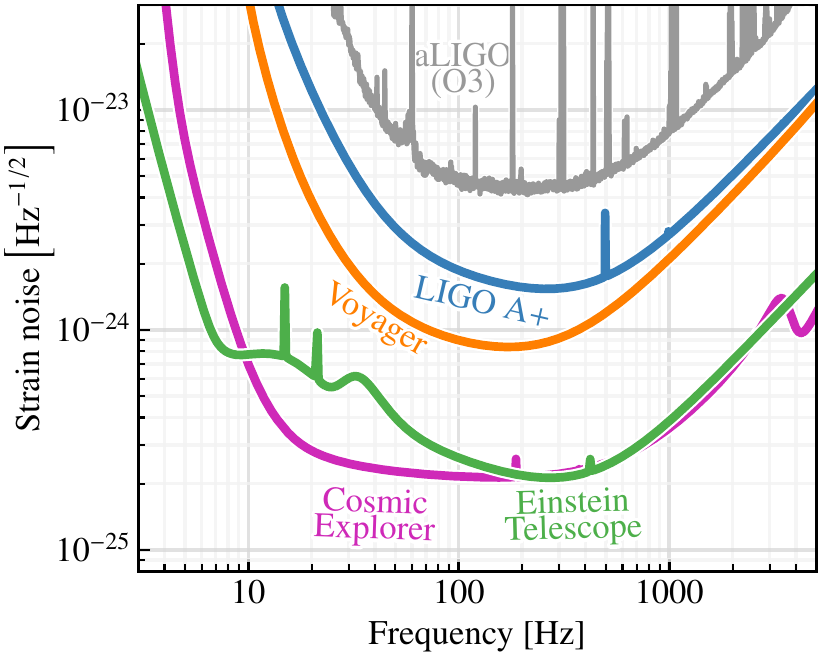}
    \caption{Observatory strain noise amplitude for optimally oriented, monochromatic GW signals. Einstein Telescope comprises three triangular detectors; the others are \textsf{L}-shaped single detectors.}
    \label{fig:strains}
\end{figure}

For both \textsf{L}-shaped and $\triangle$-shaped observatories, maxima in the antenna patterns
exist at zenith and nadir in the facility's local coordinate system.\footnote{This is true so long
as the the travel time of the light along the detector arms is negligible compared to the period of
the gravitational wave~\citep{Essick:2017wyl}.} For $\overline{F}_{\mathsf{L}}$ and
$\overline{F}_\triangle$, these maxima are $\sqrt{1/2}$ and $\sqrt{9/8}$ respectively. It is
important to note that the sensitivity of a given observatory is set not only by the antenna
pattern, but also but the detector arm length $L$ and the power spectral density $S_{x}(f)$ of the
displacement noise in each detector. Taking all these factors into account, Cosmic Explorer and
Einstein Telescope (summing over all three detectors) have nearly the same sensitivity to GW signals
in the frequency range $\SI{200}{\Hz} \lesssim f \lesssim \SI{2}{\kHz}$ (see \cref{fig:strains}).

As a consequence, we may directly compare the antenna sensitivities here for both 
\textsf{L}-shaped and $\triangle$-shaped observatories in the following, defining 
the \emph{normalized antenna power}:
\begin{equation}\label{eq:normP}
    P_{d}(\alpha,\delta;\,t) = 
    \begin{cases}
        2\,\overline{F}_{\mathsf{L}}^{2}(\alpha,\delta;\,t) &\mathrm{for}\,\,d=\mathsf{L}\,, \\
        \frac{8}{9}\,\overline{F}_{\triangle}^{2}(\alpha,\delta;\,t) &\mathrm{for}\,\,d=\triangle\,.\\
    \end{cases}
\end{equation}
The total GW signal power collected at time $t$ by a given observatory for a particular source is proportional to this normalized antenna power.

\begin{figure*}[htp]
    \centering
    \includegraphics[width=0.9\textwidth]{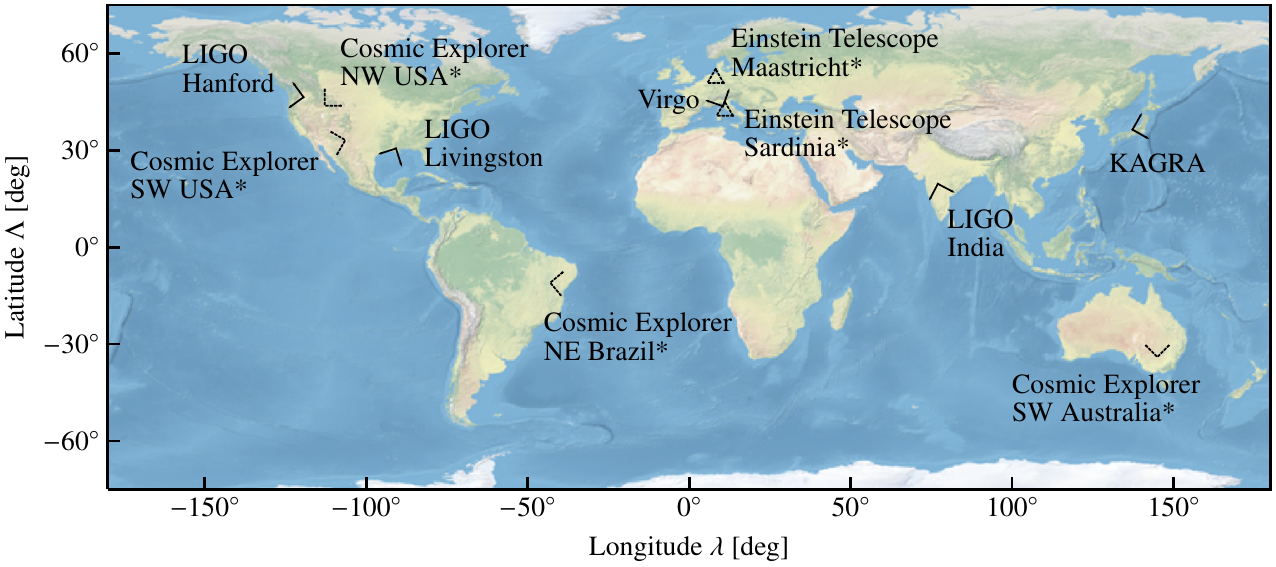}
    \caption{Location and orientation of gravitational-wave observatories. Currently operational and under-construction observatories are shown with solid lines. In this study we consider the potential observatory locations marked in dotted lines; these observatories have not been constructed. The coordinates for all sites shown in this Figure are shown in \cref{tab:det_coords}.}
    \label{fig:detectors-globe}
\end{figure*}

\begin{deluxetable*}{l l r r r r}
\tablecaption{Detector Coordinates. The latitudes $\Lambda$ and longitudes $\lambda$ of the detector vertices are given in degrees, as are the bisector angles $\gamma$ of the X and Y arms and the opening angles $\zeta$. The azimuths are defined as counterclockwise relative to due East. The coordinates for Hanford, Livingston, Virgo, and Kagra are taken from \citet{lalsuite}; those for India are taken from Bilby \citep{Ashton:2018jfp}; those for Einstein Telescope are taken from the locations of recent seismic studies \citep{Bader:2021,Allocca:2021opl}; and those for Cosmic Explorer are assumed to be the same as in \citet{Borhanian:2020ypi}.  Sites marked with asterisks are shown for illustrative purposes only. \label{tab:det_coords}}
\tablehead{
    \colhead{Facility} &
    \colhead{Code} & 
    \colhead{\phantom{0}$\Lambda$ (deg)} &
    \colhead{\phantom{00}$\lambda$ (deg)} &
    \colhead{\phantom{00}$\gamma$ (deg)} &
    \colhead{\phantom{0}$\zeta$ (deg)}
    }
\startdata
% Coords in degrees \\
% Name & Code & Lat. & Long. & X Az. & X Alt. & Y Az. & Y. Alt \\
LIGO Hanford & H & 46.455 & $-$119.408 & 170.999 & 90 \\
% LIGO Hanford   X Az.,X Alt.,Y Az.,Y. Alt, 125.000 & 0.000 & $-$144.000 & 0.000 \\
%%% Located on Native lands of the Yakima and Ichishkiin
LIGO Livingston & L & 30.563 & $-$90.774 & 242.716 & 90 \\
%$-$162.284 & 0.000 & $-$72.283 & 0.000 \\
%%% Located on Native lands of the Choctaw
Virgo & V & 43.631 & 10.504 & 115.567 & 90 \\
%70.567 & 0.000 & 160.567 & 0.000 \\
KAGRA & K & 36.412 & 137.306 & 15.396 & 90\\
%29.604 & 0.003 & 119.604 & $-$0.004 \\
LIGO India & I & 19.613 & 77.031 & 287.384 & 90 \\
%117.616 & 0.000 & 207.617 & 0.000 \\
%Einstein 1* &  \\
%Einstein 2* &  \\
%Einstein 3* &  \\
\hspace{-1.1ex}*Einstein Telescope, Meuse--Rhine & EM & 50.756 & 5.907 & 0.000 & 60 \\
\hspace{-1.1ex}*Einstein Telescope, Sardinia & ES & 40.443 & 9.457 & 0.000 & 60 \\
\hspace{-1.1ex}*Einstein Telescope, Southwestern Australia & EA & $-$34.000 & 145.000 & 90.000 & 60 \\
\hspace{-1.1ex}*Einstein Telescope, Northeastern Brazil & EB & $-$11.000 & $-$43.000 & $-$5.000 & 60 \\
\hspace{-1.1ex}*Cosmic Explorer, Northwestern USA & CN & 43.827 & $-$112.825 & 45.000 & 90 \\
%0.000 & 0.000 & 90.000 & 0.000 \\
%%% Located on Native lands of the Shoshoni
\hspace{-1.1ex}*Cosmic Explorer, Southwestern USA & CS & 33.160 & $-$106.480 & 195.000 & 90 \\
%$-$150.000 & 0.000 & $-$60.000 & 0.000 \\
%%% Located on Native lands of the Mescalero Apache
\hspace{-1.1ex}*Cosmic Explorer, Southeastern Australia & CA & $-$34.000 & 145.000 & 90.000 & 90 \\
%$-$45.000 & 0.000 & 45.000 & 0.000
%%% Located on Native lands of the Wiradjuri
\hspace{-1.1ex}*Cosmic Explorer, Northeastern Brazil & CB & $-$11.000 & $-$43.000 & $-$5.000 & 90 \\

%%% Territories of Indigenous peoples from https://native-land.ca

\enddata
\end{deluxetable*}

The above can be generalized to compute the normalized antenna power for a network of $M$ detectors 
as follows
\begin{align}
    P_{\mathcal{N},d_{1}+\ldots+d_{M}} (\alpha,\delta;t) &= \sum_{i=1}^{M}\,P_{d_{i}}(\alpha,\delta;\,t)\,,
    \label{eq:normPnet}
\end{align}
where $P_{d}(\alpha,\delta;\,t)$ is the normalized antenna power for an individual detector, as
defined in \cref{eq:normP}. In general, each term in the sum in \cref{eq:normPnet} should be
preceded by a weighting factor that describes each detector's sensitivity to the signal of 
interest. In our case, we fix all these weighting factors at unity because Einstein Telescope 
and Cosmic Explorer have nearly equal peak sensitivities from a few hundred hertz to a few 
kilohertz, which is the frequency range of interest for many Galactic science targets. In the 
case when one detector is significantly more sensitive than the others in a network, the 
overall network antenna pattern will be dominated by the single, most sensitive detector.

%%%%%%%%%%%%%%%%%%%%%%%%%%%%%%%%%%%%%%%%%%%%%%%%%%%%%%%%%%%%%%%%%%%%%%%%%%%%%%%%
Given the importance of Galactic sources to the science case for the third generation of
ground-based GW observatories, our goal is to establish how the placement and orientation of future
observatories impacts the sensitivity of the global observatory network to the Milky Way. In pursuit of
this goal, we seek here to develop figures of merit  that can be used to characterize the
directional network sensitivity to Galactic sources for both transient and continuous GW emission.

Concerning the expected Galactic source populations, we note that the prospective 
sources all concern either massive stars or (relatively nascent) 
compact objects following their explosions. As such, we expect their 
distribution across the Galaxy to closely trace that of the stellar density profile. Following 
\citet{Adams:2013ana}, we employ the profile
\begin{equation}
    \rho_\text{MW}(r,z) \propto \rme^{-r/r_{0}} \rme^{-|z|/z_{0}},
    \label{eq:gal_density}
\end{equation}
where $r$ is the radial distance (within the Galactic plane) from the Galactic center, $z$ is the
height above or below the Galactic plane, and $r_{0}= \SI{2.9}{\kpc}$ and $z_0 = \SI{95}{\pc}$ are
the characteristic scales of the distribution. In this model, the Sun is located at $(r_{\odot} =
\SI{8.7}{\kpc}, z_{\odot} = \SI{24}{\pc})$.

From \cref{eq:gal_density}, we see that the Galactic source population is expected 
to be most dense at the Galactic center, falling off as one 
moves from the Galactic plane ($z=0$), and beyond the central $\SI{3}{\kpc}$ bar. With this 
in mind, we consider scenarios concerning two different source populations:
\begin{enumerate}
    \item One concentrated at the Galactic center, and
    \item One distributed across the Galaxy according to \cref{eq:gal_density}.
\end{enumerate}

For the first case, we compute the normalized antenna power at the Galactic center 
$P_{d}(\alpha_{\mathrm{GC}},\delta_{\mathrm{GC}};t)$, considering this quantity as both (1) 
a function of sidereal time over the day, and (2) a cumulative quantity as a fraction of the 
total observing time.

In the second scenario, we numerically integrate the density profile in \cref{eq:gal_density} along
the line of sight to arrive at a weighting function $w_{\text{MW}}(\alpha,\delta)$ that indicates
the (relative) number of Galactic sources at that particular sky location. From this, we can compute 
the cumulative density function of the Galactic observations, which can be written formally as
\begin{equation}
    C_P(P_*;t) = \iint\limits_{\mathcal{S}(P_*;t)} \rmd\alpha \, \rmd\delta \, w_\text{MW}(\alpha,\delta)\,,
    \label{eq:Pcumulative_time}
\end{equation}
where $\mathcal{S}(P_*;t) = \{(\alpha,\delta) : P(\alpha,\delta;t) < P_*\}$ is the region of the sky
at time $t$ for which the antenna power falls below a threshold value $P_*$. Given $C_P(P_*;t)$ we
can read off the (time-varying) median and other quantiles of the network's antenna power pattern
for Galactic sources. Note that we can alternatively compute the cumulative density function over
all observation times,
\begin{equation}
    C_P(P_*) = \iiint\limits_{\mathcal{S}(P_*)} \rmd t \, \rmd\alpha \, \rmd\delta \,  w_\text{MW}(\alpha,\delta)\,,
    \label{eq:Pcumulative_all}
\end{equation}
with $\mathcal{S}(P_*) = \{(\alpha,\delta,t) : P(\alpha,\delta;t) < P_*\}$, from which we can
read off static quantiles characterizing the network's overall daily performance.

%One may then sum this to produce a total histogram over all source observations as a fraction of the 
%total observation time, comparable to that produced for the Galactic center case.

Given these figures of merit, we now outline how best to connect measures of their temporal behavior 
with the detectability of either transient or long-duration GW sources.

We consider first science targets that rely on the observation of a rare, short-duration burst like
a core-collapse supernova. As such an event cannot be scheduled,\footnote{if only!} it may well
occur at a time when only one observatory in the global network is online. It is, therefore,
important to characterize the observatory antenna patterns individually as a function of time, to
establish how one's ability to observe such events varies over the sidereal day. In the context of
multiple observatories, a network yielding \emph{consistent} antenna power is preferred over one with
highly variable sensitivity for transient sources.  A network that is very sensitive for, say, $\sim10\%$ of the sidereal day, but rather insensitive for the remainder of the observation time, is
not optimal for transient observation if the source occurs when the network is least sensitive.

Contrastingly, for a continuous-wave source (such as, for example, a deformed pulsar), signal power
can be accrued over a long period of time. As such, it is most instructive for these sources to
study the statistical properties of the antenna power over one sidereal day. For this purpose, we
look to the mean sensitivity of the network to establish its utility for science targets dependent
on long-duration observations of sources with continuous GW emission.

\subsection{Third-generation GW observatory networks}

With the above considerations in mind, we can compute the antenna power for a variety of
\textsf{L}-shaped and $\triangle$-shaped observatories to better understand the impact of observatory
placement and detector geometry on  detectability of Galactic sources.  Although we can (and do) consider
observatory locations chosen randomly on the globe, we will pay particular attention to a small
selection of sites, some corresponding to proposed observatory locations, and others chosen for
illustrative purposes only.  In particular, the Einstein Telescope will possibly be located either on
Sardinia or in the Meuse--Rhine Euroregion~\citep{Amann:2020jgo}, and so we consider sites that are
close to these locations.  The location of Cosmic Explorer is less certain, but is likely to be
somewhere in the Western United States; we therefore choose two illustrative sites, one in the
Northwest and another in the Southwest, which have been used in previous third-generation studies 
(see, e.g.,~\citet{Borhanian:2020ypi}). We also consider two illustrative observatory sites 
located outside Europe and North America: one in Southeastern Australia, and another in Northeastern Brazil.

In \cref{fig:detectors-globe}, we show the placement of the observatory sites that we have
considered, along with the sites of current observatories.  For each featured site, the specific
parameter set characterizing the detector geometry (latitude $\Lambda$, longitude $\lambda$,
opening angle $\zeta$, and arm bisector angle $\gamma$, as previously introduced) is listed in
\cref{tab:det_coords}.  In the following sections we will consider these observatory 
sites, in addition to 30,000 sites placed randomly across the globe, in one-,
two-, and three-node network configurations. The reason to consider networks of varying size is that
third-generation observatories may not begin operating at the same time.  Additionally,
observatories may go offline for maintenance or upgrades, or because of environmental disturbances;
experience from the third observing run of the second-generation GW observatory network shows that,
given three observatories participating in the run, all three are simultaneously online just under
half the time~\citep{LIGO:2021ppb}.

\section{Single Observatory Placement}\label{sec:results-onedet}

We now consider how various third-generation GW observatories in isolation observe the Galaxy (i.e., a
one-observatory network).  We consider sources at the Galactic center in
Section~\ref{sec:results-onedet-GC}, and a source population distributed across the Milky Way in
Section~\ref{sec:results-onedet-galplane}.

\subsection{Galactic center sources}\label{sec:results-onedet-GC}
\begin{figure*}[htp]
    \centering
     \includegraphics[width=0.9\textwidth]{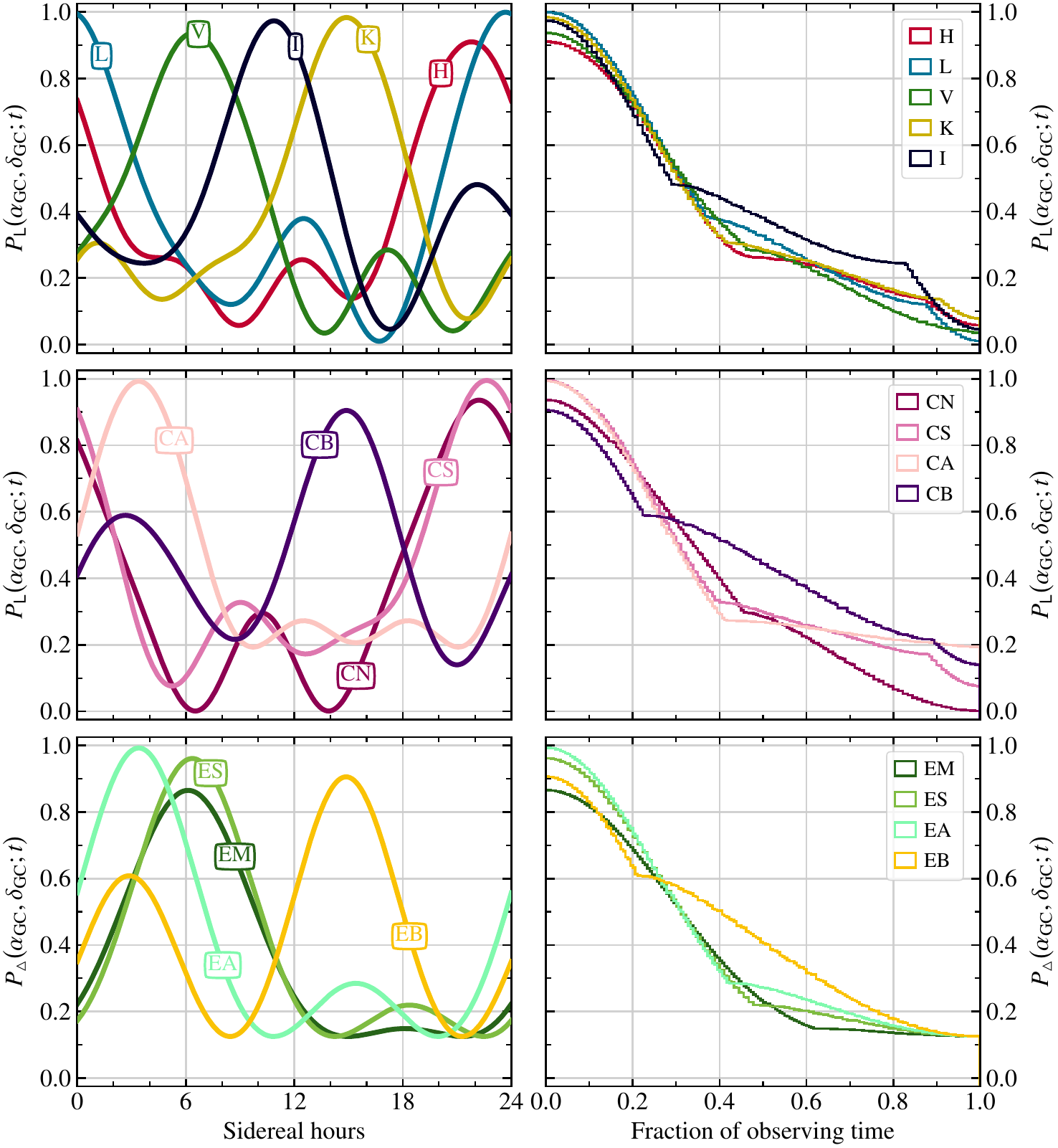}
     \caption{Top panels: for a source at located at the Galactic center, the normalized,
polarization-averaged antenna power of the existing ($\mathsf{L}$-shaped) ground-based GW observatories
shown as (left) a function of sidereal hour over one sidereal day, and (right) a
cumulative histogram over fractional daily observing time. The same is
shown for (middle panels) the locations for ($\mathsf{L}$-shaped) Cosmic Explorer, and
(bottom panels) the locations for the ($\triangle$-shaped) Einstein Telescope. The locations and
orientations for all observatories featured are given in \cref{tab:det_coords}.
\label{fig:normP_GC_onedet}}
\end{figure*}

In \cref{fig:normP_GC_onedet}, we show the normalized polarization-averaged antenna power for
sources at the Galactic center as a function of both sidereal hours and cumulative fraction of daily
observing time, for (1) the current global network of GW observatories, (2) four potential locations for
Cosmic Explorer, and (3) four potential locations for the Einstein Telescope.

For a single \textsf{L}-shaped Cosmic Explorer type observatory, we find that both the Southeastern 
Australia (\texttt{CA}) and Southwestern US (\texttt{CS}) locations reach near optimal sensitivity 
for sources directed toward the Galactic center for around an hour per sidereal day. For 
proposed locations in Northwestern US (\texttt{CN}) and Northeastern Brazil (\texttt{CB}), the maximal 
antenna power is around 0.9. Looking instead however at the antenna power as a function 
of fractional observation time rather than sidereal hour, we see that the \texttt{CB} 
location yields a higher antenna power for longer, with $P^{\mathtt{CB}}_{\mathsf{L}}\lesssim0.45$ 
for over half the observation time, in comparison with $P_{\mathsf{L}}\lesssim0.25$ for over 
half the observation time for the proposed \texttt{CA}, \texttt{CN}, and \texttt{CS}  
locations.

Comparatively, for a single $\triangle$-shaped Einstein Telescope type observatory, the Australian site
(\texttt{EA}) reaches near optimal sensitivity similar to the \texttt{CA} site.  We note there is
far less variation between the two proposed sites in Meuse--Rhine (\texttt{EM}) and Sardinia
(\texttt{ES}) than for the four sites considered for Cosmic Explorer type observatories, in both maximal
achieved antenna power and temporal coincidence of these peaks. This, in truth, is to be expected
given the geographical proximity of the \texttt{EM} and \texttt{ES} sites. While it can be seen that
the maximal antenna power achievable is slightly larger at the \texttt{ES} site than at the \texttt{EM}
site (roughly a $\sim$10\% effect), we note that this improvement is only seen for around $20\%$ of
the total observation time. For the remaining $80\%$ of the sidereal day, the antenna powers yielded
at each site are comparable. As in the \textsf{L}-shaped case, the Brazilian site yields a higher
antenna power for a larger fraction of the day in comparison to the other sites.

\begin{figure*}[htbp]
    \centering
    \includegraphics[width=0.9\textwidth]{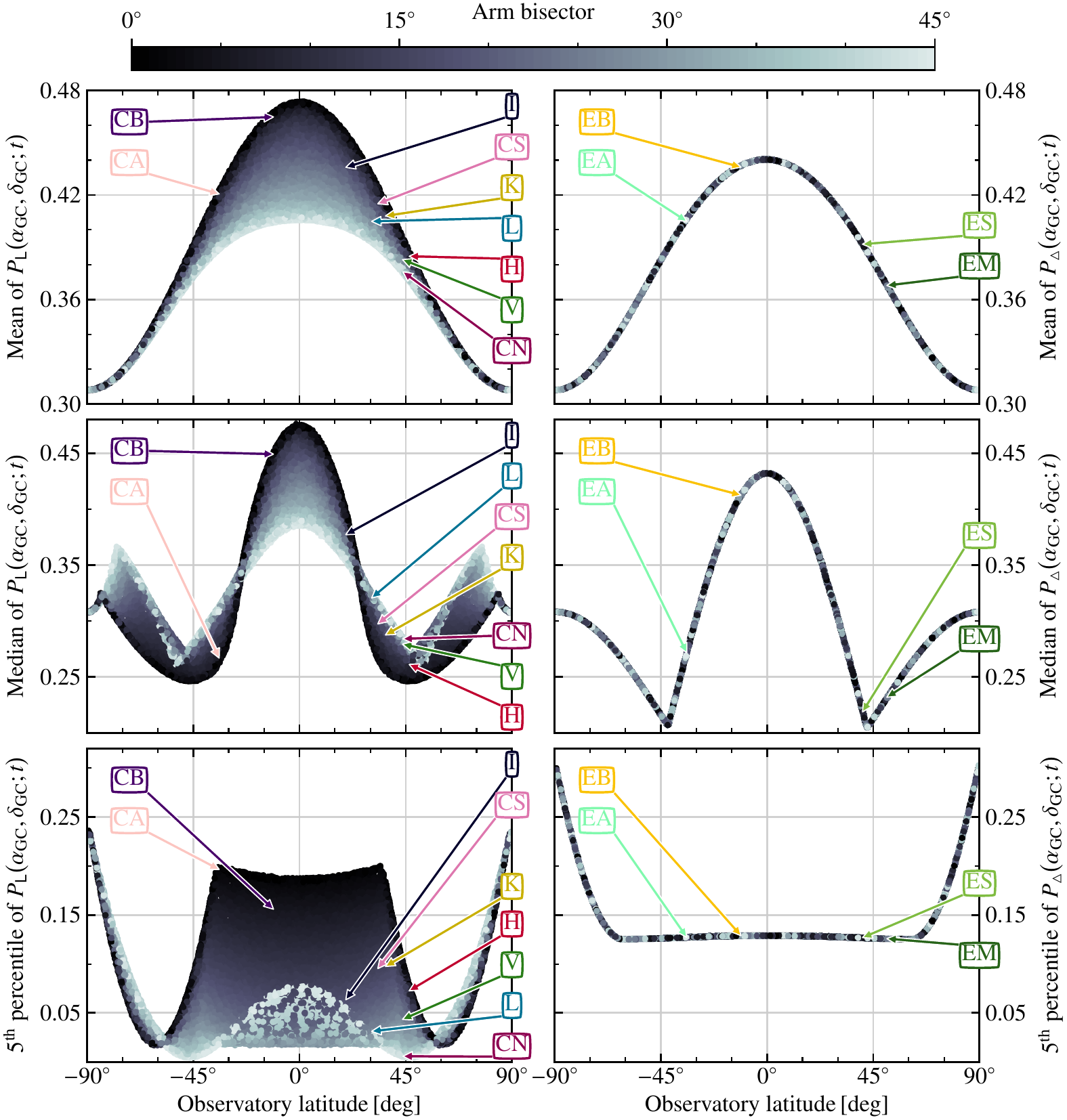}
    \caption{Left column: given 30,000 randomly placed 
    $\mathsf{L}$-shaped observatories observing a source at the Galactic center, the mean, 
    median, and fifth percentile of the polarization-averaged 
    antenna power over one sidereal day (top, middle, and 
    bottom panels, respectively) are shown. In each panel, the latitude and 
    arm bisector angle of currently existing observatories are annotated, along with
    the Cosmic Explorer locations considered. Right column: as for the left column, but in the 
    context of 30,000 randomly placed $\triangle$-shaped observatories (each comprising 
    three detectors combined). In each panel, the Einstein Telescope sites considered
    are annotated.} 
    \label{fig:normP_GC_latitude}
\end{figure*}

For the purpose of improving detection prospects for long-duration and/or continuous GW sources, it
is useful to look at the statistical properties of the daily antenna power for both the observatory
sites in \cref{tab:det_coords} and for hypothetical sites placed randomly in latitude, longitude,
and detector orientation. In \cref{fig:normP_GC_latitude}, we show the mean, median, and fifth
percentile of the antenna power for 30,000 randomly placed observatories, both for
\textsf{L}-shaped (Cosmic Explorer type) and $\triangle$-shaped (Einstein Telescope type)
observatories. 

An immediate takeaway is that -- for both types of observatory -- the mean and median antenna powers
over one sidereal day (to sources located at the Galactic center) are maximized with equatorial
observatory placement.

For \textsf{L}-shaped detectors, we find the mean
$P_{\mathsf{L}}(\alpha_{\mathrm{GC}},\delta_{\mathrm{GC}};t)$ is augmented by between $30\%$ and
$50\%$ for equatorial versus polar placement, with maximal improvements yielded with the arm
bisector oriented along due East $\gamma=\SI{45}{\degree}$ (versus configurations for which the
detector arms lie along the compass directional axes, $\gamma=\SI{0}{\degree}$). 

For the median $P_{\mathsf{L}}(\alpha_{\mathrm{GC}},\delta_{\mathrm{GC}};t)$, maximal antenna power
is once more found through equatorial observatory placement and orientation with detector arm bisector along due East, yielding a factor of two increase over a similar detector placed at latitudes
$\pm\SI{45}{\degree}$.

Finally for the fifth percentile of
$P_{\mathsf{L}}(\alpha_{\mathrm{GC}},\delta_{\mathrm{GC}};t)$, we find maximal values this time for
observatories located either directly at the poles (irrespective of orientation), or latitudes
satisfying $|\Lambda|\le\SI{35}{\degree}$ (oriented with $\gamma=\SI{0}{\degree}$).  For the latter
scenario, we note that detector orientation is very important: almost an order of magnitude
variation is seen between $\gamma=\SI{0}{\degree}$ and $\gamma=\SI{45}{\degree}$ (due to the
quadrupole nature of the antenna power pattern), which has a resounding impact on prospects for
transient sources. 

Of the Cosmic Explorer sites considered in \cref{tab:det_coords}, we note that the \texttt{CN}
location is quite suboptimal for sources located around the Galactic center, particularly when
compared to the alternatives. Overall, \texttt{CB} performs best across all antenna power metrics
here, followed by \texttt{CA} (which exhibits a poor median, but a fair mean and near maximal fifth percentile), and \texttt{CS} (with a fair median, mean, and fifth percentile).

For $\triangle$-shaped observatories, we find similar trends across the mean and median antenna
powers. For equatorial sites, the mean $P_{\triangle}(\alpha_{\mathrm{GC}},\delta_{\mathrm{GC}};t)$
is increased by up to $\sim42\%$ over their polar counterparts, with no impact from observatory
orientation, since the overall antenna pattern of the triple detector configuration
(\cref{eq:Fbar_triangle}) is azimuthally symmetric in the observatory's local basis.  The median
$P_{\triangle}(\alpha_{\mathrm{GC}},\delta_{\mathrm{GC}};t)$ is maximized through equatorial
placement, with up to a factor of two improvement yielded over other sites. As for \textsf{L}-shaped
observatories , we find here that the median is minimized for sites at latitudes
$\Lambda\simeq\pm\SI{45}{\degree}$.  For the fifth percentile of
$P_{\triangle}(\alpha_{\mathrm{GC}},\delta_{\mathrm{GC}};t)$, antenna power is maximized for
observatories placed directly at the poles (roughly a factor of two increase), while minimized at
latitudes satisfying $|\Lambda|\le\SI{60}{\degree}$. While this may seem catastrophic in light of the previous optimization for equatorial observatories, we do note that the minimum 
fifth percentile for
$P_{\triangle}(\alpha_{\mathrm{GC}},\delta_{\mathrm{GC}};t)\sim0.13$, which is comparatively better
than for the majority of \textsf{L}-shaped observatories. This is a consequence of the fact that unlike
the antenna pattern of a single detector, the antenna pattern of the triple detector configuration
has no nulls, attaining a minimum of 1/8 the maximum power.

Of the Einstein Telescope sites considered in \cref{tab:det_coords}, we see that the Brazilian site
\texttt{EB} outperforms the others in terms of mean, median, and fifth percentile.
Between the two European sites, there is little difference in performance. While the mean antenna
power for \texttt{ES} shows a roughly $10\%$ improvement over \texttt{EM}, \texttt{EM} in turn
exhibits a median antenna power of around $10\%$ better than for \texttt{ES}, and they both have nearly identical fifth percentiles in antenna power. 

\begin{figure*}[p]
    \centering
    \includegraphics[width=0.9\textwidth]{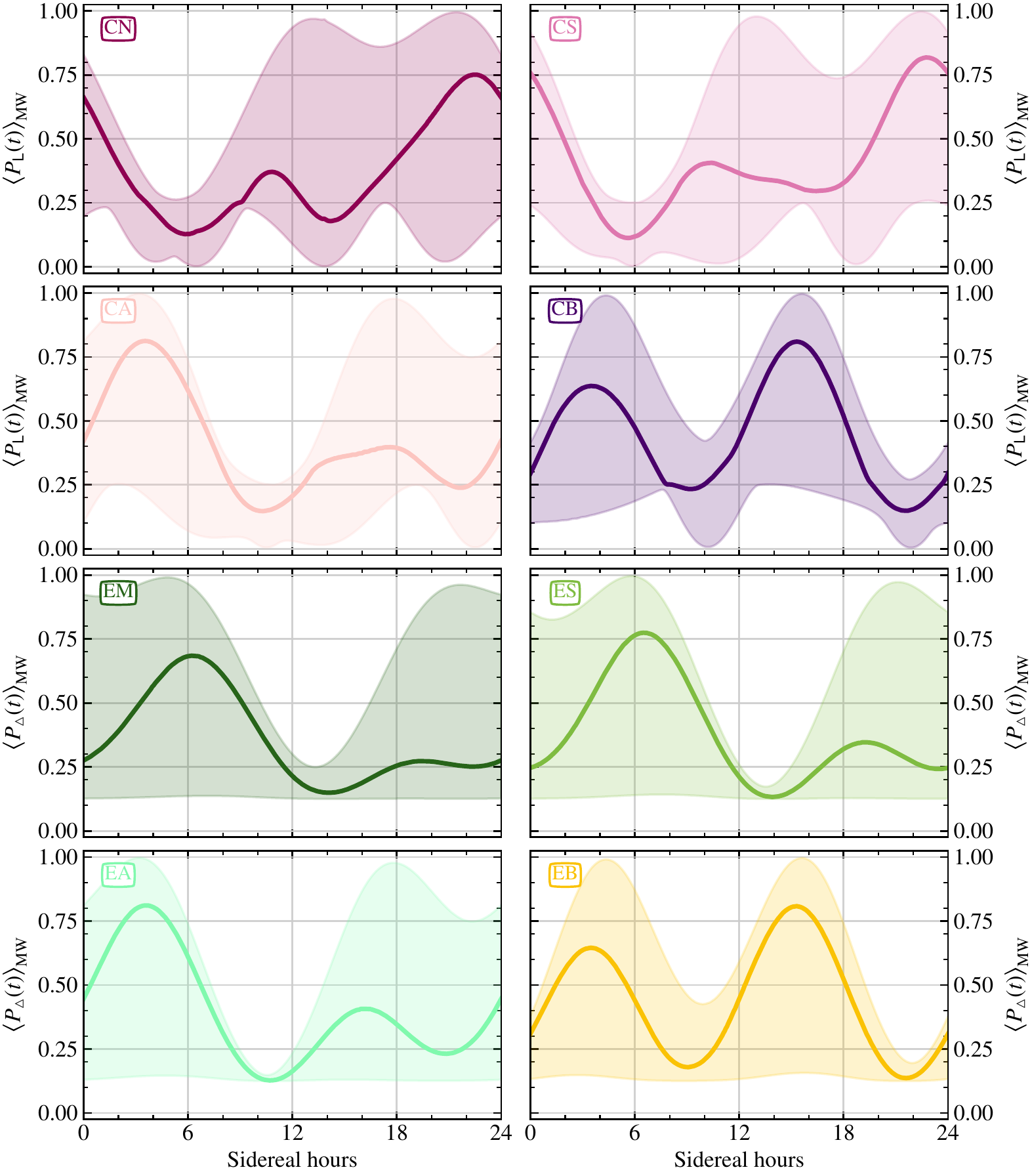}
    \caption{
     Averaged across an astrophysically motivated distribution of Galactic sources (see
\cref{eq:gal_density}), the normalized, polarization-averaged antenna power over one sidereal day
for potential Cosmic Explorer (top and middle panels) and Einstein Telescope (bottom panels)
observatory locations. In each panel, the solid line shows the median antenna power across the
considered source population, while the banding spans between the 5\textsuperscript{th} and
95\textsuperscript{th} percentiles.
    \label{fig:normP_gal_onedet}}
\end{figure*}

\subsection{Galactic plane average}
\label{sec:results-onedet-galplane}
While the density of GW sources is expected to be concentrated about the Galactic center, we do
indeed expect the source population to be distributed across the Galaxy, following the stellar
density. As this has an impact on our results from the previous section, we recompute the antenna
power in the context of an astrophysically motivated distribution of Galactic sources. 

In \cref{fig:normP_gal_onedet}, we show the normalized polarization-averaged antenna power over the
sidereal day for the Cosmic Explorer and Einstein Telescope sites considered in
\cref{tab:det_coords}, averaged over the anticipated population of Galactic GW sources as in
\cref{eq:Pcumulative_time}.  Unsurprisingly, there is spread in the antenna power over the sidereal
day, unlike for the Galactic center case in \cref{fig:normP_GC_onedet}. We note that the median behavior
seen here does loosely trace the antenna power for a single source at the Galactic center. This is
to be expected as, based on our imposed model for source density, most of the sources are
concentrated in the central regions of the Galaxy. Keeping in mind, however, that the Galactic plane
traverses most of the celestial sky over one sidereal day, it is natural that averaging antenna
power over a distribution of Galactic sources would result in some spread -- as observed.

\begin{figure*}[tp]
    \centering
    \includegraphics[width=0.9\textwidth]{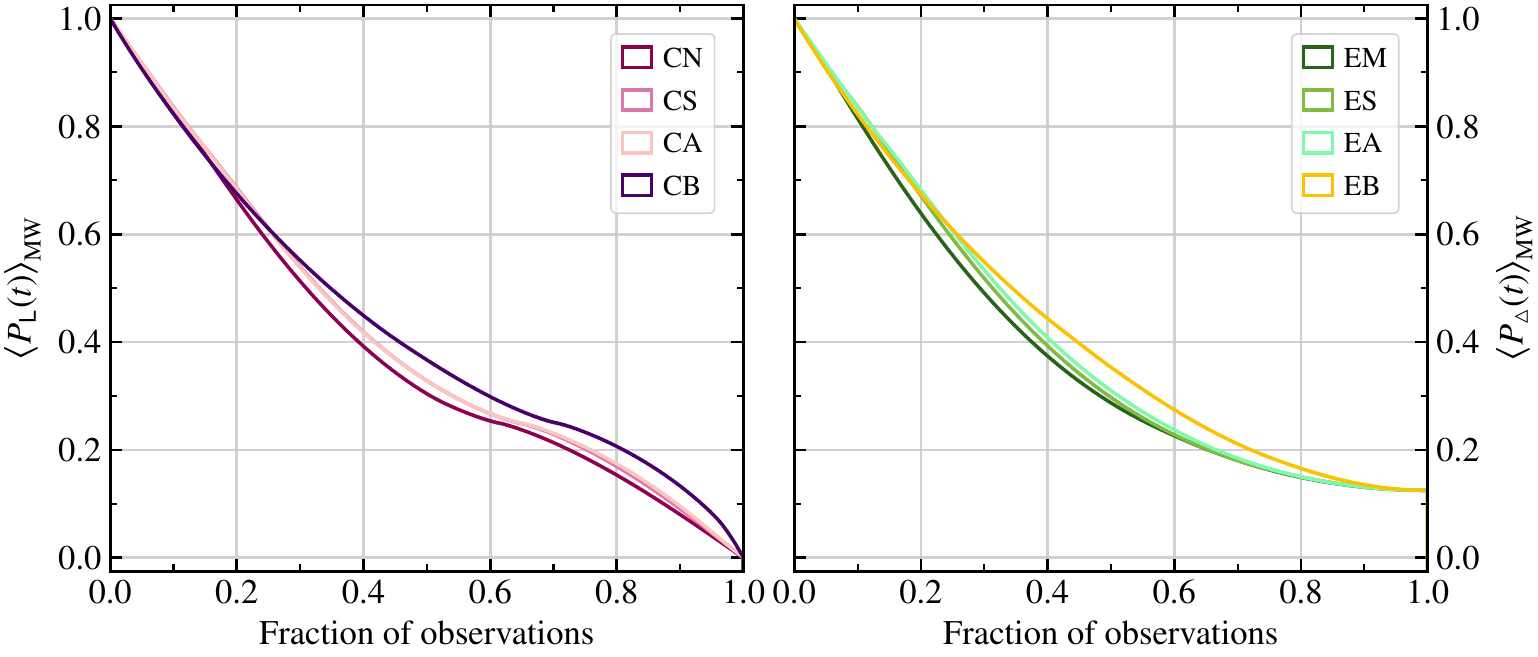}
    \caption{
     Averaged across an astrophysically motivated distribution of Galactic sources 
     (see \cref{eq:gal_density}), the normalized, polarization-averaged antenna power 
     over one sidereal day for (left panel) possible locations of a 
     Cosmic Explorer observatory, and (right panel) possible locations of an Einstein 
     Telescope observatory.}
    \label{fig:normP_gal_onedet_hist}
\end{figure*}

In \cref{fig:normP_gal_onedet_hist}, we show the normalized polarization-averaged antenna power for
the same astrophysically motivated distribution of Galactic sources, again for the
locations for Cosmic Explorer and the Einstein Telescope considered, but this time as a function of the fractional
observing time as in \cref{eq:Pcumulative_all}.  Comparing these results to the Galactic center case
(as seen in the middle right and bottom right panels of \cref{fig:normP_GC_onedet}), we see that any
discontinuities in the cumulative behavior of the antenna power (caused, ostensibly, by the point
nature of the single sky position considered) are -- as expected -- smoothed out as we average over
the Galaxy. We note that any small differences between the two Einstein Telescope sites in Europe
have all but vanished for the Galactic average scenario, suggesting that (should equatorial
placement not be feasible for the Einstein Telescope) both the \texttt{EM} and \texttt{ES} sites
would perform equivalently in the context of Galactic science, particularly when placed in a network
with other observatories. A similar, but not identical, statement can be made about the Cosmic Explorer
sites considered. We see that the antenna power (as a fraction of the observing time) is
indistinguishable between the \texttt{CN} and \texttt{CS} sites, with only a minor improvement seen
for the \texttt{CA} site. Finally, the \texttt{CB} site edges out the other three sites for the
purpose of observing the Milky Way, though not by a great amount. This suggests that the marked
gains in antenna power seen from equatorial observatory placement are muted somewhat when observing
across the Galaxy rather than just the Galactic center.

\section{Observatory Network Placement}\label{sec:results-detnetwork}
While optimizing the placement of a single observatory is an important consideration, 
it is of interest to investigate how further gains to the Galactic science case can 
be achieved through optimal placement of a network, given one or more fixed observatory 
sites. 

\subsection{Two-observatory network}\label{subsec:twodet}
Although we showed in Section~\ref{sec:results-onedet} that equatorial 
placement of both \textsf{L}-shaped and $\triangle$-shaped observatories maximizes the antenna 
power for sources at the Galactic center, realistically it is expected that the Einstein 
Telescope will be built near either the \texttt{EM} or \texttt{ES} locations (leaving open the possibility
that additional observatories could be built at \texttt{EA} or \texttt{EB} at some later time). As we 
have seen that, in the context of a Galactic distribution of GW sources, there is 
negligible difference in performance between the two sites, we choose one at random to 
fix the $\triangle$-shaped observatory in our considered networks. 

We consider herein the performance of a two-observatory network (\texttt{EM+Cx}), 
comprising (1) one $\triangle$-shaped observatory located in Meuse--Rhine,
and (2) one arbitrarily placed \textsf{L}-shaped observatory, in the context of source populations 
concentrated in the Galactic center, and distributed across the Milky Way.
    
\subsubsection{Galactic center}
\begin{figure*}[tph]
    \centering
     \includegraphics[width=0.9\textwidth]{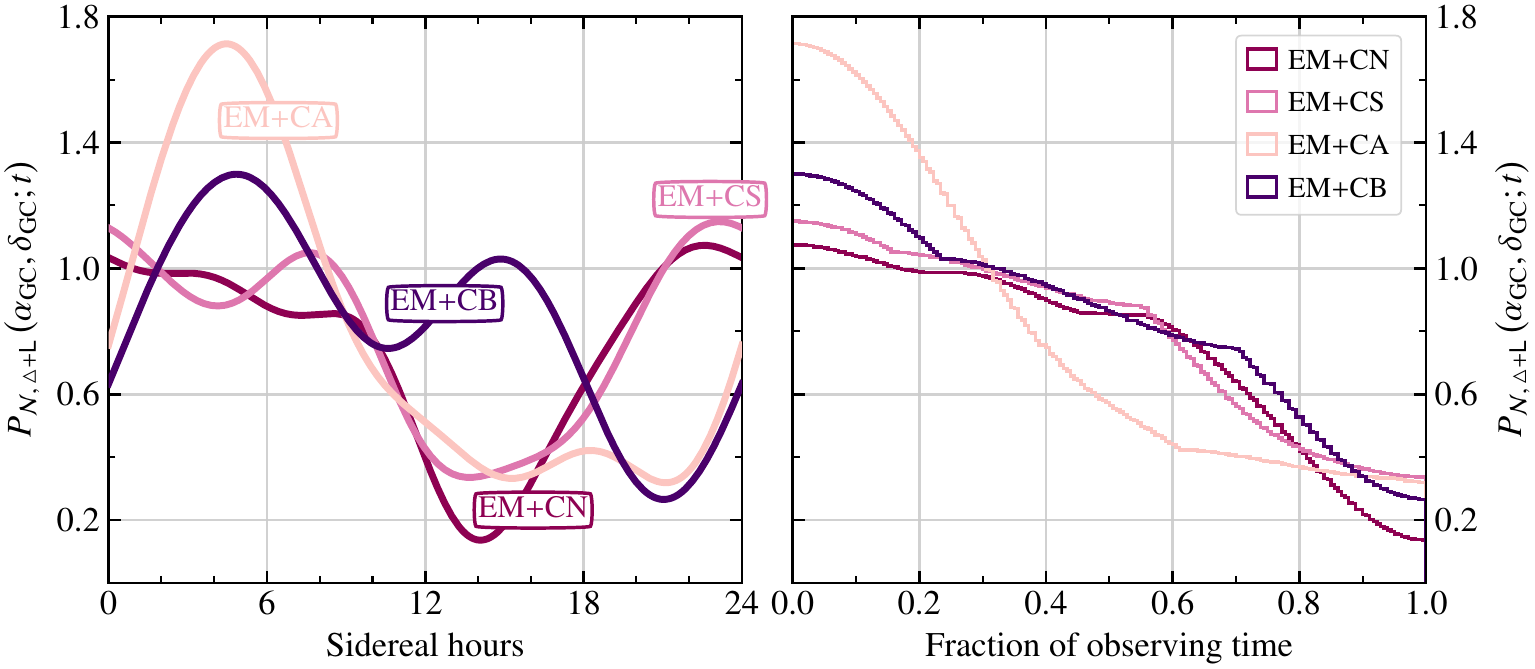}
     \caption{For a source at the Galactic center, the normalized, 
     polarization-averaged antenna power as (left) a function of sidereal hour, and 
     (right) a cumulative histogram over fractional daily observing time, in the context 
     of a two-observatory network (\texttt{EM+Cx}; comprising 
     (1) a $\triangle$-shaped observatory located in Meuse--Rhine, and 
     (2) a \textsf{L}-shaped observatory located at one of the four 
     Cosmic Explorer sites).}
    \label{fig:normP_GC_twodet}
\end{figure*}

In \cref{fig:normP_GC_twodet}, we show the normalized, polarization-averaged antenna 
power for a two-observatory network (\texttt{EM+Cx}; comprising one $\triangle$-shaped observatory 
located in Meuse--Rhine, and one \textsf{L}-shaped observatory placed at one of the four Cosmic 
Explorer sites) observing the Galactic center as a function of both sidereal hours and cumulative 
fraction of daily observing time.

Looking at the antenna power as a function of sidereal time, we note (at first glance) that 
the network \texttt{EM+CA} (of the four Cosmic Explorer sites) is most sensitive to 
the Galactic center, while network \texttt{EM+CN} is the least sensitive. Taking into account, 
however, the cumulative behavior as a fraction of observing time, we see that such a simplistic 
statement is misleading. While network \texttt{EM+CA} does indeed have a peak antenna power 
greater than networks with \textsf{L}-shaped detectors at one of the other three Cosmic 
Explorer sites, the network is only more sensitive than the others for $\sim30\%$ of the 
observing time. For more than half the observing time, network \texttt{EM+CA} is in fact 
significantly less sensitive than its counterparts. Comparatively, while network 
\texttt{EM+CN} is least sensitive for $\sim15\%$ of the day, the antenna power yielded is 
comparable to that for network \texttt{EM+CS} for the remainder of the observation time. Of 
the four networks considered here, \texttt{EM+CB} does slightly better over the entire 
observational period, with slightly improved ($\sim20\%$ effect) antenna power for the best 
$\sim20\%$ of the sidereal day, while maintaining a similar sensitivity to networks 
\texttt{EM+CN} and \texttt{EM+CS} for the remainder of the time.

\begin{figure}[tph]
    \centering
    \includegraphics[width=\columnwidth]{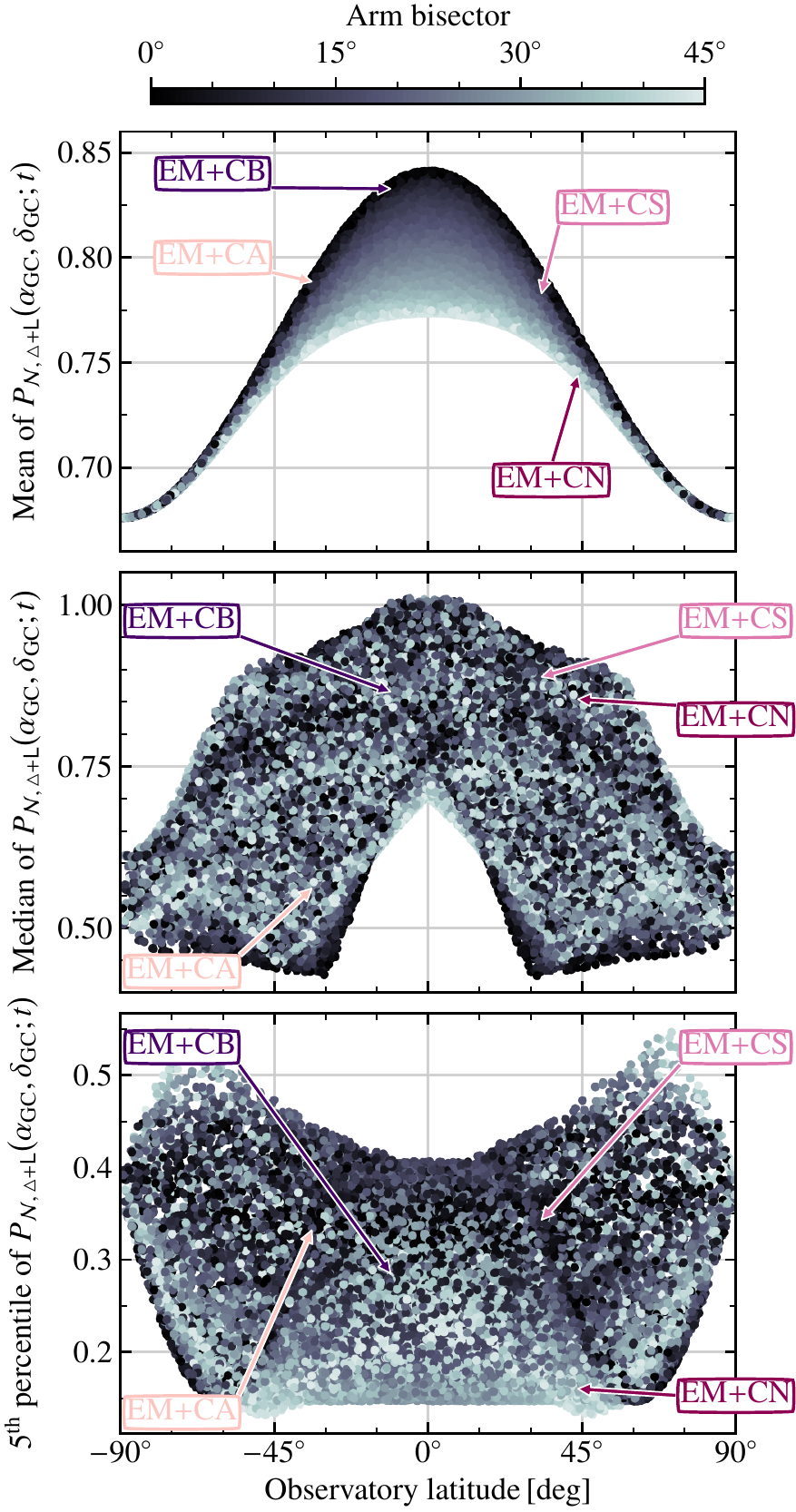}
    \caption{
      Given a two-observatory network (\texttt{EM+Cx}; comprising (1) a $\triangle$-shaped observatory located in Meuse--Rhine, and (2) one of 30,000 randomly placed \textsf{L}-shaped 
      observatories) observing a source at the Galactic center, the mean, 
    median, and fifth percentile of the polarization-averaged 
    antenna power over one sidereal day (top, middle, and 
    bottom panels, respectively) are shown. In each panel, annotations indicate two-observatory 
    networks in which the \textsf{L}-shaped observatory is placed in one of the Cosmic 
    Explorer sites given in \cref{tab:det_coords}.}
    \label{fig:normP_GC_twodet_stats}
\end{figure}

To better understand the impact of observatory location and orientation on network performance, we 
show in \cref{fig:normP_GC_twodet_stats} the statistical behavior of the antenna power over 
one sidereal day for a two-observatory network \texttt{EM+Cx}, comprising one $\triangle$-shaped 
observatory in Meuse--Rhine, and one \textsf{L}-shaped observatory arbitrarily placed on the globe.

As for the single-observatory case, we see that the mean antenna power is maximized when the 
\textsf{L}-shaped observatory is located equatorially with arm bisector 
angle $\gamma=\SI{0}{\degree}$. 
We find that equatorial placement alone yields at minimum a $\sim15\%$ improvement over polar 
placement, while an up to an additional $\sim7\%$ increase in mean antenna power can be gained 
for equatorial observatories by orienting the \textsf{L}-shaped detector with the arm bisector 
along due East. Of the four Cosmic Explorer sites considered, we see that placing the 
second observatory at \texttt{CB} 
results in an almost maximal mean network antenna power. Comparatively, we note that both 
\texttt{EM+CA} and \texttt{EM+CS} show similar (and fair) mean antenna power, while network 
\texttt{EM+CN} site performs worst of all four networks considered, and quite poorly. 

For the median and fifth percentile of the antenna power, the trends with observatory
location and orientation are less straightforward, but broadly follow similar trends with latitude 
as seen for the one-observatory case. Again, the median antenna power is maximized with equatorial 
observatory placement, but there is no clear trend with arm bisector angle. We also see that the peak in 
median antenna power is spread out in latitude, with as little as a $\sim15\%$ reduction seen for some 
observatories in latitudes $\Lambda\in[-60^{\circ},\,60^{\circ}]$.

Similarly for the fifth percentile of network antenna power, the top-hat shaped distribution 
seen for the single \textsf{L}-shaped observatory case in \cref{fig:normP_GC_latitude} is smeared out 
across latitude and arm bisector angle. There is variation across 
observatories, with as much as a factor of five difference between the ``worse-case'' 
and ``best-case'' scenarios, but there is a far weaker correlation in this behavior seen 
with position and orientation.

While this may at first seem counterintuitive, reflection on the results for a single 
\textsf{L}-shaped observatory as a function of sidereal time yields some clarity. We saw 
previously that equatorial placement (and, less so, detector orientation with $\gamma=\SI{0}{\degree}$) 
maximizes single-observatory antenna power. We see this reflected in the mean behavior of the 
network antenna power, but not for the median and fifth percentile. The reason 
for this is that the variation of the network antenna power over one sidereal day is dependent 
on not only on the peaks in antenna power for the single observatories, but on the relative 
sensitivities of the observatories at any given point in time. For example, a large mean 
network antenna power can arise both from a consistently sensitive network, and a network very 
sensitive at one part of the day but less so for the remainder of the observation time. As we have 
seen, the location and orientation of a single observatory impact not only the amplitude of the 
antenna power peaks and troughs, but also control the time of day at which they occur. 
Optimizing a two-observatory network for Galactic observations, therefore, requires intentional placement 
of the second observatory relative to the first in the network, to ensure their peaks in antenna 
sensitivity are complementary over the sidereal day, rather than overlapping in time. 

\subsubsection{Galactic plane average}
\begin{figure*}[tph]
    \centering
    \includegraphics[width=0.9\textwidth]{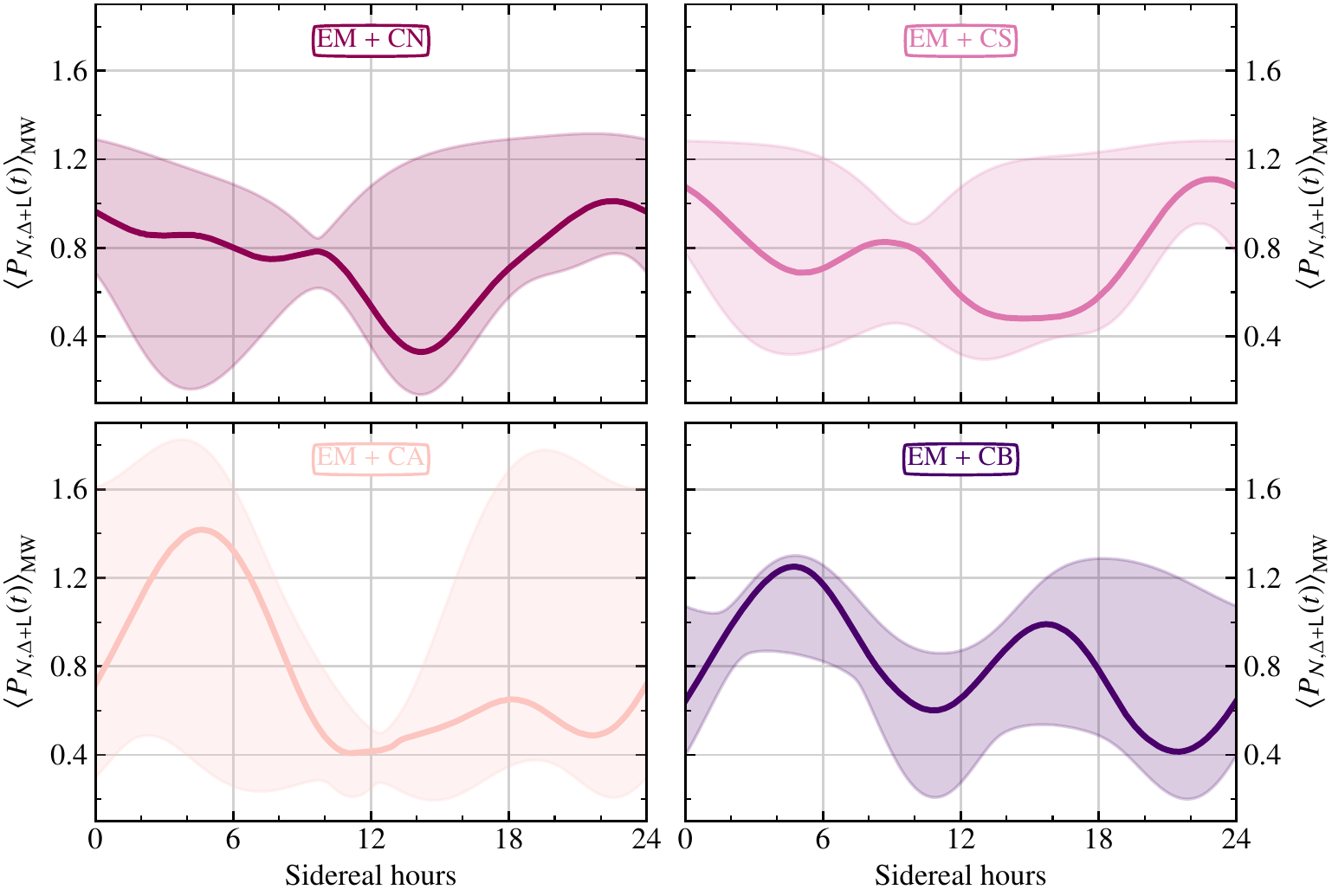}
    \caption{
    Averaged across an astrophysicallyvmotivated distribution of Galactic sources 
     (see \cref{eq:gal_density}), the normalized, polarization-averaged antenna power 
     over one sidereal day for a two-observatory network (\texttt{EM+Cx}; comprising (1) a 
     $\triangle$-shaped observatory located in Meuse--Rhine, 
     and (2) a \textsf{L}-shaped observatory located at one of the four 
     Cosmic Explorer sites), is shown. As in \cref{fig:normP_gal_onedet}, 
     the solid line shows the median antenna power across the considered source population, 
     while the banding spans between the 5\textsuperscript{th} and 95\textsuperscript{th} percentiles.
     }
    \label{fig:normP_gal_twodet}
\end{figure*}

Turning our attention once more to a population of sources distributed across the Galaxy, 
in \cref{fig:normP_gal_twodet} we show the normalized 
polarization-averaged antenna power over the sidereal day for a two-observatory network 
(\texttt{EM+Cx}), comprising one $\triangle$-shaped observatory in Meuse--Rhine, and 
one \textsf{L}-shaped observatory placed at one of the four Cosmic Explorer sites considered.

Similarly to the single-observatory scenario (see \cref{fig:normP_gal_onedet}), we note a spread in the network antenna power over the sidereal day, unlike for the Galactic center case in \cref{fig:normP_GC_twodet}. Once more, the median 
behavior loosely traces the results for the Galactic center. The largest variation occurs for the 
\texttt{EM+CA} network, particularly around times where the network antenna power peaks. 
Contrastingly, far less variation is seen for the \texttt{EM+CN/CS/CB} networks which, 
while experiencing smaller amplitude peaks in network antenna power, appear to exhibit a more 
consistent network sensitivity across the sidereal day. 

\begin{figure}[tph]
    \centering
    \includegraphics[width=\columnwidth]{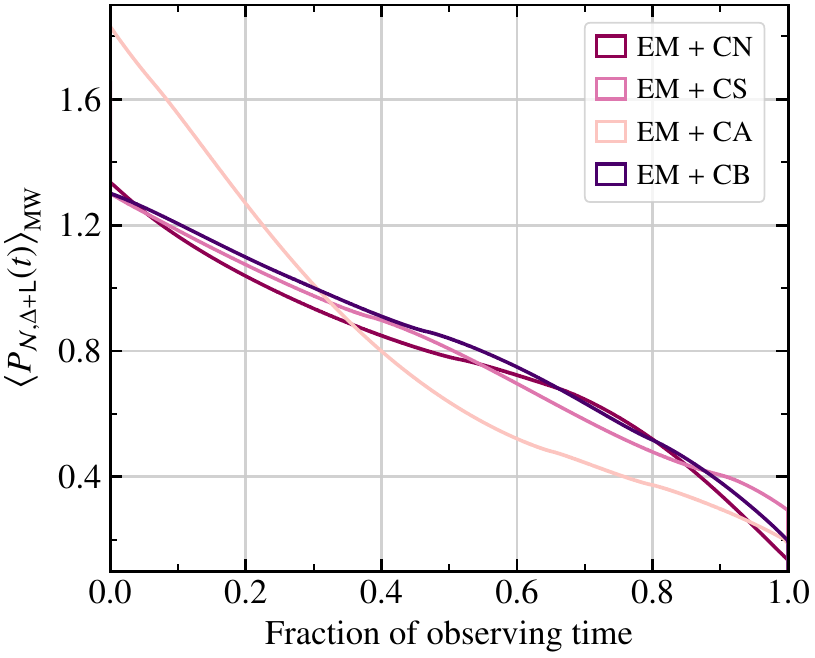}
    \caption{Averaged across an astrophysically motivated distribution of Galactic sources 
     (see \cref{eq:gal_density}), the normalized, polarization-averaged antenna power 
     over one sidereal day for a two-observatory network (\texttt{EM+Cx}; comprising (1) a 
     $\triangle$-shaped observatory located in Meuse--Rhine, and (2) a \textsf{L}-shaped 
     observatory located at one of the four Cosmic Explorer sites considered).}
    \label{fig:normP_gal_twodet_hist}
\end{figure}

The previous conclusions are confirmed in \cref{fig:normP_gal_twodet_hist}, where we show the cumulative 
histogram of network antenna power (averaged over the Galactic source population). 
While the \texttt{EM+CA} network is up to $\sim30\%$ more sensitive than \texttt{EM+CN/CS/CB} for 
$\sim30\%$ of the observational time, it is also less sensitive (by almost $\sim50\%$ at worst) than 
the other networks for over $50\%$ of the sidereal day. We note that the relative 
performance of the \texttt{EM+CN/CS/CB} networks is very similar, with near negligible differences 
when compared over the total observing time. As such, they are likely more suitable for targeting 
transient observations than the \texttt{EM+CA} network due to their more consistent network antenna 
power. In turn, the \texttt{EM+CA} network may be preferable for continuous-wave searches due to 
the long-duration observation times involved ($\tau \gg 1$ sidereal day), which ensure the high 
network antenna power during that $\sim30\%$ of the sidereal day can consistently be used to 
the observer's advantage.

\subsection{Three-observatory network}
Akin to the aforementioned observatory placement scenario outlined for the Einstein Telescope, it 
is expected that at least one of the Cosmic Explorer observatories will be 
built in North America (i.e. either the \texttt{CN} or \texttt{CS} site). As we 
have seen in the previous Section, the performance of the \texttt{CN} and 
\texttt{CS} sites are similar in the context of a two-observatory network 
\texttt{EM+Cx}, meaning that either may be used as a node in the network near-interchangeably.
As such, we thus  
consider a three-observatory network (\texttt{EM+CS+Cx}), comprising 
(1) one $\triangle$-shaped observatory located in Meuse--Rhine, 
(2) one \textsf{L}-shaped observatory placed at the \texttt{CS} site, and 
(3) a second \textsf{L}-shaped observatory 
placed arbitrarily. As previously, we investigate the performance of the three-observatory network 
in the context of two source populations: one concentrated at the Galactic center, and one 
distributed across the Galaxy.

\subsubsection{Galactic center}

\begin{figure*}[tph]
    \centering
     \includegraphics[width=0.9\textwidth]{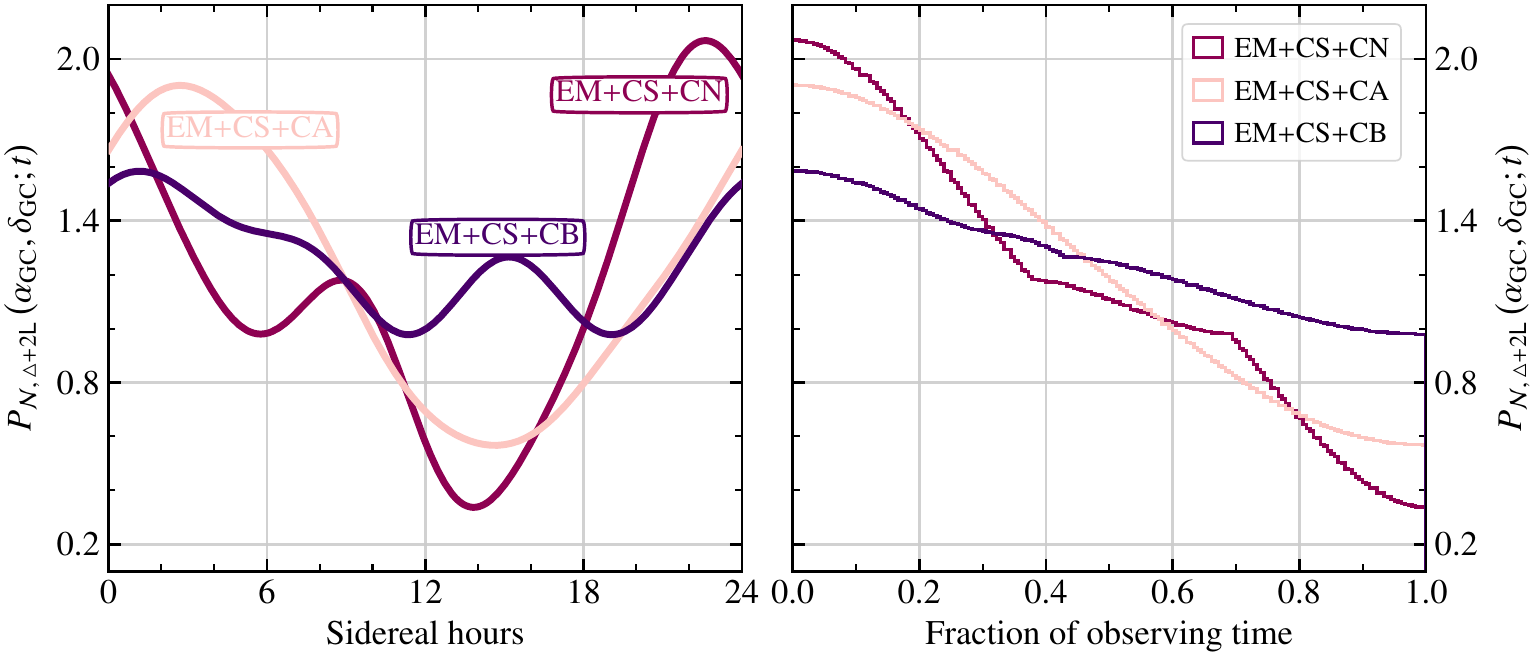}
     \caption{For a source at the Galactic center, the normalized, 
     polarization-averaged antenna power as (left) a function of sidereal hour, and 
     (right) a cumulative histogram over fractional daily observing time, in the context 
     of a three-observatory network (\texttt{EM+CS+Cx}; comprising (1) one 
     $\triangle$-shaped observatory located in Meuse--Rhine,
     (2) one \textsf{L}-shaped observatory located in the Southwestern United States, and
     (3) a second \textsf{L}-shaped observatory located at one of the other Cosmic Explorer 
     sites).
     }
    \label{fig:normP_GC_threedet}
\end{figure*}

In \cref{fig:normP_GC_threedet}, we show the normalized, polarization-averaged antenna 
power for the three-observatory network (\texttt{EM+CS+Cx}; comprising (1) one $\triangle$-shaped 
observatory located in Meuse--Rhine, (2) one North American \textsf{L}-shaped 
observatory placed at  the \texttt{CS} site, and (3) a second \textsf{L}-shaped 
observatory placed at one of the 
other three Cosmic Explorer sites) observing the Galactic center as a function of sidereal 
hour.

Of the three networks considered, we see that the \texttt{EM+CS+CN} reaches the largest peak 
antenna power, followed closely by \texttt{EM+CS+CA}. While the peak antenna power 
for \texttt{EM+CS+CB} is lower than both \texttt{EM+CS+CN/CA}, the network exhibits far less 
variation over the sidereal day. These initial ``first-glance'' suppositions are backed up by 
the cumulative behavior of the network antenna power over the full observing time. While the 
peak network antenna power for \texttt{EM+CS+CN} (\texttt{EM+CS+CA}) is $\sim28\%$ 
($\sim18\%$) greater than that for \texttt{EM+CS+CB}, the network as a whole is only more 
sensitive for $\sim30\%$ ($\sim45\%$) of the total observing time. \texttt{EM+CS+CB}, however, exhibits remarkable stability in network antenna power over the sidereal day, 
with the lowest network power well over $60\%$ of the peak. Comparing this to \texttt{EM+CS+CN}, 
the minimum network power is just $\sim17\%$ of the peak. Similarly, the minimum network 
power for \texttt{EM+CS+CA} is $\sim31\%$ of its peak. This would suggest that the 
\texttt{EM+CS+CB} network is preferable for observation of transient events (which 
cannot be scheduled), while \texttt{EM+CS+CN/CA} may be favorable for continuous observations. Of 
\texttt{EM+CS+CN} and \texttt{EM+CS+CA}, the latter is more suitable as it confers less variability 
in sensitivity, while still exhibiting a relatively large peak network power. 

\begin{figure}[tph]
    \centering
    \includegraphics[width=\columnwidth]{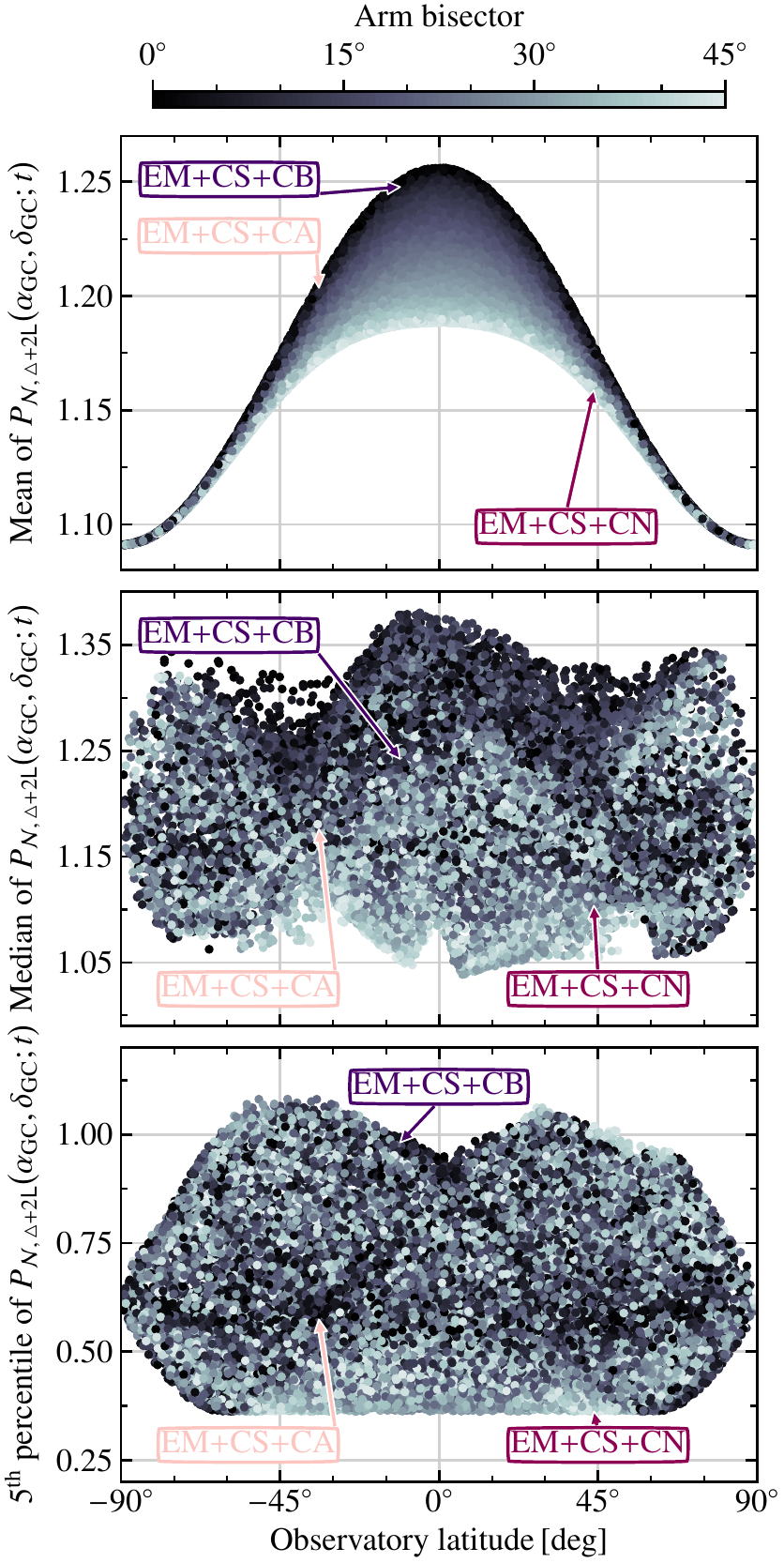}
    \caption{
    Given a three-observatory network (\texttt{EM+CS+Cx}; comprising 
    (1) a $\triangle$-shaped observatory located in Meuse--Rhine, 
    (2) one \textsf{L}-shaped observatory located in the Southwestern USA, and 
    (3) one of 30,000 randomly placed \textsf{L}-shaped observatories) observing a 
    source at the Galactic center, the mean, 
    median, and fifth percentile of the polarization-averaged 
    antenna power over one sidereal day (top, middle, and 
    bottom panels, respectively) are shown. In each panel, the three-observatory 
    networks in which the second \textsf{L}-shaped observatory is placed at one of the 
    other Cosmic Explorer sites are annotated.
    }
    \label{fig:normP_GC_threedet_stats}
\end{figure}

To once more better understand the impact of observatory location and orientation on network performance, 
we show in \cref{fig:normP_GC_threedet_stats} the statistical behavior of the antenna power over 
one sidereal day for a three-observatory network \texttt{EM+CS+Cx}, where the third observatory is
randomly placed on the globe.
We see that, as for the one- and two-observatory cases, the mean network power is 
maximized when the third observatory is placed at equatorial latitudes (and with arm bisector 
$\gamma=\SI{0}{\degree}$ for \textsf{L}-shaped observatories with latitudes 
$\Lambda \in [-\SI{60}{\degree}, \SI{60}{\degree}]$), which can yield up to a $\sim15\%$ 
improvement over polar counterparts. Of the networks considered here, \texttt{EM+CS+CB} performs best 
for Galactic sources, again followed by \texttt{EM+CS+CA} and then \texttt{EM+CS+CN}. 

With that being said, the relationship between placement of the third observatory in the network 
and the 5\textsuperscript{th} and 50\textsuperscript{th} percentiles of network antenna power is 
more complex. The distribution of network power with time is 
dependent not only on the observatory latitude (which determines the single-observatory peak antenna 
power), but also on the observatory longitude (which determines the time at which the peak antenna 
power occurs). Regardless, we do note that variation across the 30,000 randomly placed observatories 
is (somewhat) 
maximized by equatorial placement, but the longitude must be chosen such that the peak of 
antenna power does not overlay that of the other observatory nodes. To conclude, we note that (of the 
networks we considered) \texttt{EM+CS+CB} is preferable for Galactic observations.

\subsubsection{Galactic plane average}
\begin{figure}[tph]
    \centering
    \includegraphics[width=\columnwidth]{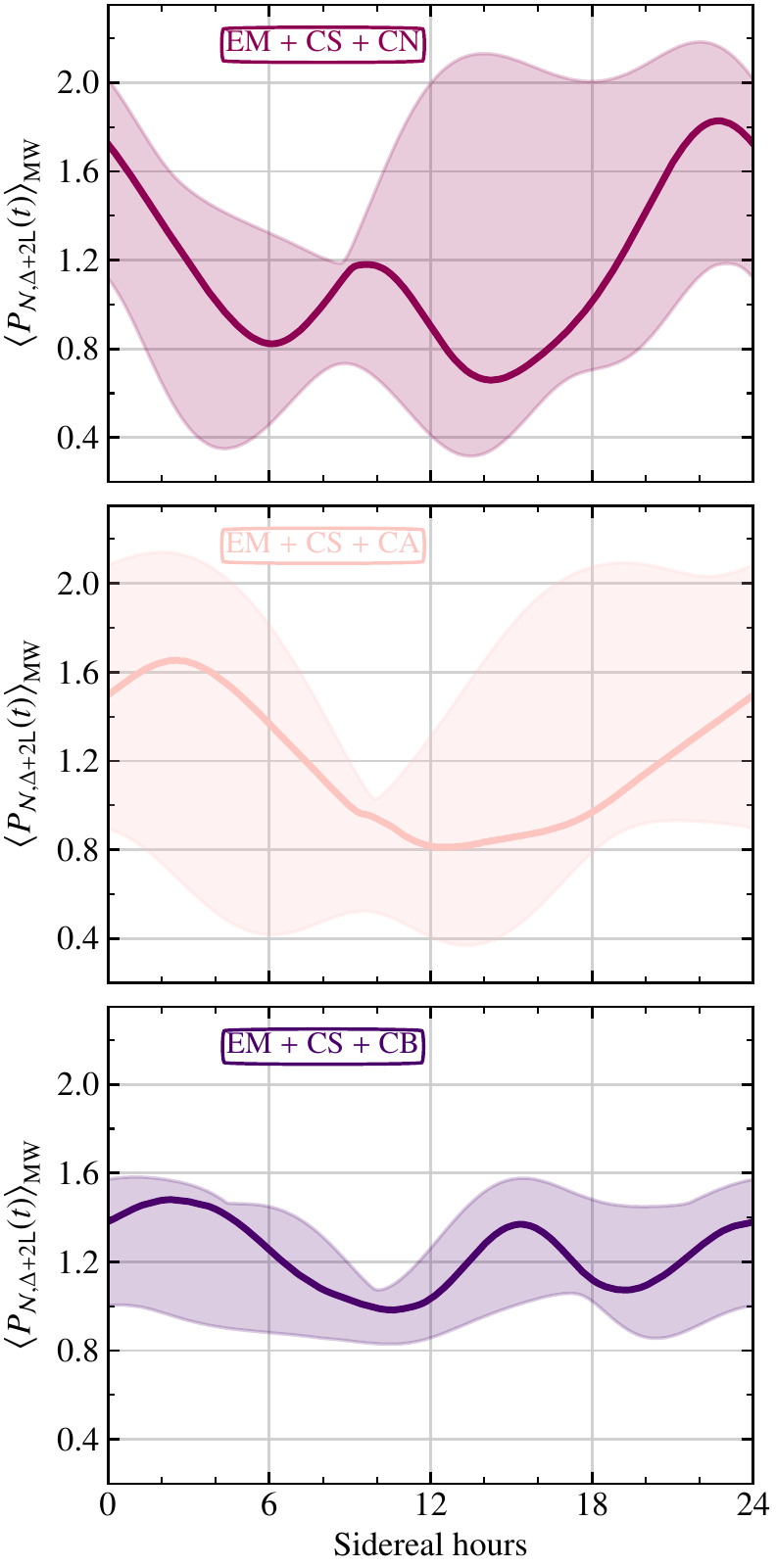}
    \caption{
    Averaged across an astrophysically motivated distribution of Galactic sources 
     (see \cref{eq:gal_density}), the normalized, polarization-averaged antenna power 
     over one sidereal day for a three-observatory network (\texttt{EM+CS+Cx}) comprising (1) one 
     $\triangle$-shaped observatory located in Meuse--Rhine,
     (2) one \textsf{L}-shaped observatory located in the southwest USA, 
     and (3) a second \textsf{L}-shaped observatory located in the (top panel) Northwestern USA, (middle panel) Southeastern Australia, or (bottom panel) Northeastern Brazil. 
     As in \cref{fig:normP_gal_onedet} and \cref{fig:normP_gal_twodet}, 
     the solid line shows the median antenna power across the considered source population, 
     while the banding spans between the 5\textsuperscript{th} and 95\textsuperscript{th} percentiles.
}
    \label{fig:normP_gal_threedet}
\end{figure}

Considering one final time a population of sources distributed across the Galaxy, we 
show in \cref{fig:normP_gal_threedet} the normalized 
polarization-averaged antenna power over the sidereal day for a three-observatory network 
\texttt{EM+CS+Cx}, comprising one $\triangle$-shaped observatory in Meuse--Rhine,  
one North American \textsf{L}-shaped observatory placed at \texttt{CS}, 
and a second \textsf{L}-shaped observatory placed at one of \texttt{CN}, \texttt{CA}, or \texttt{CB}.

\begin{figure}[tph]
    \centering
    \includegraphics[width=\columnwidth]{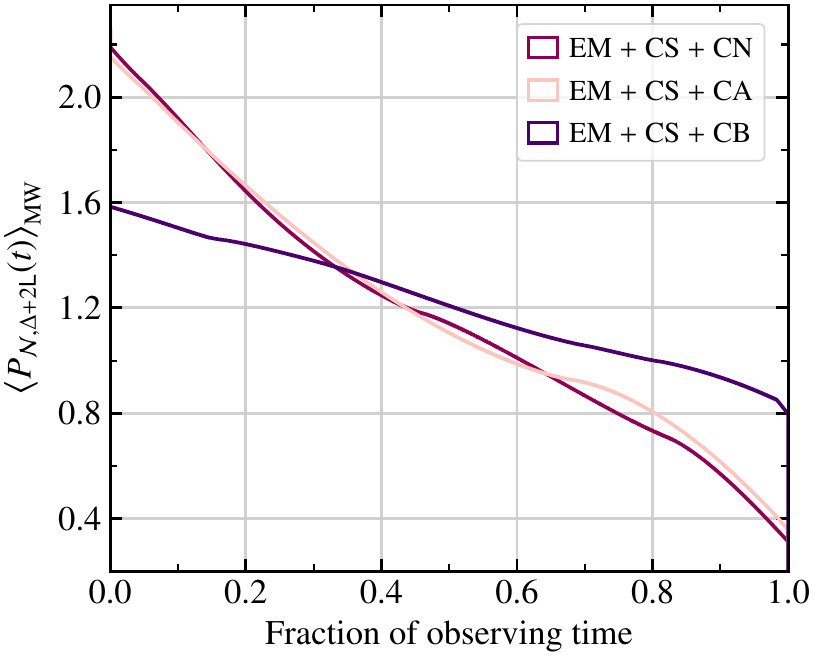}
    \caption{
    Averaged across an astrophysically motivated distribution of Galactic sources 
     (see \cref{eq:gal_density}), the normalized, polarization-averaged antenna power 
     over one sidereal day for a three-observatory network (\texttt{EM+CS+Cx}; comprising 
     (1) a $\triangle$-shaped observatory located in Meuse--Rhine,
     (2) one \textsf{L}-shaped observatory located in the Southwestern USA, and 
     (3) a second \textsf{L}-shaped observatory located in one of the other Cosmic 
     Explorer sites).}
    \label{fig:normP_gal_threedet_hist}
\end{figure}

Similarly to the Galactic center case, we see that networks \texttt{EM+CS+CN/CA} have higher peaks in 
antenna power (albeit with greater variability across the source distribution), while \texttt{EM+CS+CB} 
exhibits more sensitivity (with less variability across the Galaxy) over the sidereal day. In 
\cref{fig:normP_gal_threedet_hist}, which shows the cumulative 
histogram of network antenna power as a fraction of observing time, we see these hunches are confirmed.

Averaging over the Galactic source population, the differences in performance between 
\texttt{EM+CS+CN} and \texttt{EM+CS+CA} have all but vanished, with only minute improvements 
exhibited by \texttt{EM+CS+CA} for the least sensitive $30\%$ of the observation time. 
Comparatively, \texttt{EM+CS+CB} has a peak antenna power $25\%$ lower than 
\texttt{EM+CS+CN/CA}, but it is more sensitive than both for $\sim65\%$ of the observation 
time. Addressing now variation in antenna power over the sidereal day, we see once more 
that \texttt{EM+CS+CB} exhibits far greater consistency with just a factor of two between its 
peak and minimum sensitivity. By contrast, for both \texttt{EM+CS+CN/CA}, the minimum 
antenna power is just $\sim15\%$ of the peak. Reiterating the conclusions of the previous 
Section, we suggest that \texttt{EM+CS+CB} is the optimal three-observatory network for Galactic 
observations of those considered here for both the consistent sensitivity and fair peak antenna 
power.

\section{Discussion}\label{sec:discussion}

While previous studies have focused on source-specific figures of merit such as 
detection volumes, polarization reconstruction, localization, and inference of compact binary 
parameters, we focus here solely on a source-independent figure of merit -- the network antenna 
power. Although not specific to any particular source, this choice is indeed purposeful. The Milky Way is abundant with 
prospective GW sources, yielding signals of widely varying (and often uncertain) morphologies. As 
such, we aim here to quantify solely the prospective improvement in source detectability owing 
to augmented antenna power, rather than tuning to source properties that -- at this time -- are 
rather ambiguous. 

It is important to acknowledge the limitations to this approach. 
First, we note that while antenna power may well be correlated to the total signal energy collected by the network, the relationship between the total energy and the actual signal-to-noise ratio recovered by a data analysis pipeline -- particularly 
for poorly modeled sources -- is not well understood.\footnote{This is in contrast to the situation for well-modeled sources, for which all of the signal energy can be recovered via a matched filter in the ideal case~\citep{WainsteinZubakov}.} As such, we do caution 
against direct one-to-one extrapolation from augmented network antenna power to expected 
detection volume within the Galaxy. While the two will no doubt be related, we remind the 
reader that excess-power searches -- even those employing coherent network analysis -- 
recover only a fraction of the signal-to-noise ratio that would be extracted by other methods 
(such as matched filtering) not applicable to the sources considered here. While not the focus of this paper, gains in detectability 
comparable to 
those presented here may well be possible through development of targeted analysis 
methods (see, e.g.,~\citet{Thrane:2014bma,Millhouse:2018dgi,Cornish:2020dwh,Gossan:Hall:inprep}).

Second, it is important to ensure that in optimizing 
the third-generation of GW observatories for Galactic science, one examines the effect on 
other science goals. Source localization, which in the current generation of observatories 
is done primarily through triangulation 
based on differences in detection times between sites, is dependent on the distances between 
the observatories along the line of sight to 
the GW source in question. As a result, accuracy across the sky is improved by maximizing the 
area of the polygon formed by the network nodes. For the two- and three-observatory networks 
considered here, the relevant metrics to maximize would be the interobservatory distance and 
area of the triangle by the observatories, respectively~\citep{Schutz:2011tw,Raffai:2013yt}.

Relating this back to the study herein, we see that the 
optimum placement is likely to 
bolster the accuracy of source localization rather than hinder it. The reason for this is 
that, given the expectation that at least two of the third-generation observatories will 
be built in the Northern Hemisphere (with the Einstein Telescope in Europe, and 
Cosmic Explorer in North America), the directive to place a third observatory at 
near-equatorial latitudes naturally increases the triangle subtended by the network. 
Furthermore, placement in the Southern Hemisphere at either the Brazilian or 
Australian site would maximize the potential for source localization while simultaneously 
improving prospects for Galactic science. 

Third, we also note that the number of observatories online at any given moment can vary. 
During the most recent observing run for Advanced LIGO and Advanced Virgo (O3), while 
the single-observatory duty cycles topped $75\%$, the two-observatory and three-observatory duty cycles 
were just $62\%$ and $47\%$, respectively~\citep{aLIGO:2020wna,LIGO:2021ppb}. Given this, we 
reassert the importance of careful placement of network nodes with respect 
to each other. As seen in Section~\ref{sec:results-detnetwork}, colocation of observatories 
can result in a network that is very sensitive to Galactic sources for some small fraction 
of the day, while being decidedly suboptimal for the remainder of the observation time. While this 
was noted previously in the context of network utility for observation of continuous-wave 
versus transient sources, the issue of duty cycle further complicates matters. Consequently, 
it is not only relative placement of observatories in the global network that 
ought to be considered for improving antenna sensitivity to the Galactic plane, but also 
the local time zone to anticipate detector downtime from both commissioning and increased 
environmental noise. By taking this into account (in a similar manner to, 
for example,~\citet{Szolgyen:2017hfh} in the context of the already built Advanced LIGO, 
Advanced Virgo, and KAGRA observatories), one can -- in principle -- place network nodes 
in anticipation of nonoptimal duty cycles, and plan for an observatory network where the 
antenna sensitivity does not drop below some predefined threshold. While such 
considerations are beyond the scope of this study, we emphasize the importance of 
quantifying this effect, and aim to tackle it in future work. 

Considering now the impact on all-sky sensitivity, we note that while placement to 
improve antenna power for Galactic sources does come at a cost to the 
overall network isotropy, this effect is not so marked. As stated  
in~\citet{Raffai:2013yt}, while the geometry of the network will change the overall shape 
of the network antenna pattern (i.e., improved sensitivity in one direction results in 
reduced sensitivity elsewhere), the all-sky integral remains constant for different 
configurations of the detectors within the network. On the presumption of a homogeneous and 
isotropic distribution of compact object binaries within the local universe, the overall 
reach of the observatory network (as characterized by detection volume;~\citet{Schutz:2011tw}) 
is expected to be unchanged by opting for a network geometry that improves sensitivity 
along the Galactic plane. 

Building upon this point, we address the potential impact on prospects for observing 
electromagnetic counterparts to GW transients. While the Galactic plane is certainly 
electromagnetically bright, it traces out but a tiny sliver of the 
celestial sky. Per the discussion above, improving antenna sensitivity for Galactic sources 
will not reduce the detection volume averaged over the sky. As it is expected that the  
canonical 1.4--1.4\,$M_{\odot}$ binary neutron star systems will be observable out to 
$z\gtrsim4$ with both the Einstein Telescope and Cosmic Explorer~\citep{Bailes:2021tot}, the 
very slight directional preference for the Galactic plane is likely to have minimal consequences 
on the observed population of compact object binaries over the course of an observing run.  

We further note that a subset of ultracompact object binaries will be observable by 
space-based GW observatories long before they enter the target frequency band for 
ground-based detectors. 
This means that not only will the accuracy of localization will be improved, but additionally 
the advance warning will permit pointing of telescopes well ahead of merger -- enabling 
longer integration times and, consequently, improved observational prospects. To 
conclude on this point, extragalactic multimessenger astronomy will not be 
negatively impacted by placing ground-based GW observatories with Galactic optimization in mind. 
 
Finally, we highlight a crucial consideration: the impact of the 
observatory on the land, environment, and nearby communities. As stated   
in the context of Cosmic Explorer~\citep{Evans:2021gyd}, the proposed third-generation 
GW observatories should be built and operated with the 
ongoing consent of the local communities. This tenet must be central to the development, 
construction, and operation of any and all GW observatories, and prioritized by the 
collaborative teams leading such efforts from the outset.

\section{Conclusion}\label{sec:conc}
In this study, we explore whether the prospects for observation of Galactic GW sources 
can be improved through placement of the potential 
third-generation GW observatories comprising a future global 
network. To this end, we employ the antenna power pattern of the network as an appropriate figure of merit, 
considering an 
astrophysically motivated distribution of GW sources throughout the Milky Way. As 
network antenna sensitivity to a particular sky location will vary with time over the 
day (owing to the Earth's rotation), we evaluate the average behavior and 
statistical variations of the antenna power over one sidereal day, for the purpose of 
commenting on how detection prospects are expected to be impacted for both transient and 
continuous-wave sources. 

For a single observatory, we find that equatorial placement most benefits the 
observation of Galactic sources, with an up to $\sim50\%$ ($42\%$) improvement in the 
mean antenna power yielded over hypothetical \textsf{L}-shaped ($\triangle$-shaped) 
polar observatories. We note that observatory orientation has a marked impact on antenna 
sensitivity for an \textsf{L}-shaped observatory with latitudes 
$\Lambda \in [\SI{-60}{\degree},\,\SI{60}{\degree}]$, with maximization occurring when the 
arm bisector is aligned along any of the local cardinal directions. For polar \textsf{L}-shaped 
observatories and $\triangle$-shaped observatories at any latitude, orientation has 
a negligible impact on the mean antenna power at the Galactic center. Of the proposed 
sites, the Brazilian site (\texttt{CB}/\texttt{EB}) performs optimally for both types 
of observatory considered, with the Australian site (\texttt{CA}/\texttt{EA}) also being 
favorable. Considering only the sites in the Northern 
Hemisphere, we see that \texttt{CS} offers improved sensitivity over \texttt{CN} for an 
\textsf{L}-shaped observatory in North America, while the performances at either \texttt{ES} or 
\texttt{EM} for a European $\triangle$-shaped observatory are comparable. 

In the context of network placement, we see similar trends in mean antenna power, 
but note a key difference: while equatorial placement improves peak power for a single 
observatory, it is important to consider the \emph{relative} placement of observatories, as well 
as their expected duty cycle, to thoroughly understand the statistical behavior of the network 
with time. 
At all latitudes for two- and three-observatory networks, we see that the longitude of the 
additional observatory has a significant impact, which is due to its effect on the time of 
day at which the antenna power peaks. As noted previously, for a consistent network sensitivity over the sidereal day it is important to place the
observatories such that the peak antenna power for each network node occurs at different times 
of day. If the 
placement of observatories is such that their peak antenna powers temporally overlap, the network 
will be particularly sensitive at that one time, but will be far less sensitive for the remainder 
of the day. Further, the local time zones at the different observatories should be considered, such that commissioning and/or downtime due to environmental noise are  
minimized during the period of day at which the observatory is most sensitive to the Galactic 
plane (see, e.g.,~\citet{Szolgyen:2017hfh}).

Considering the performance of two- and three-observatory networks, this study primarily focuses on
the first two observatory sites being in Europe and North America, and not on the 
more generic case of optimal placement of network nodes. For instance, in the case of a 
three-observatory network, there is only a $\sim15\%$ difference 
in mean antenna power between the best and worst case scenarios. Similarly for the median 
antenna power, the worst case scenario is just $\sim20\%$ lower than for optimal placement. 
For the fifth percentile, however, there is more variation, with an improvement 
of up to a factor of four possible through optimal versus suboptimal placement.

The case for Galactic multimessenger astrophysics is rich and diverse, with much 
as-yet-untapped potential to probe the Milky Way in GWs. The construction of a 
network of third-generation ground-based GW observatories offers such an 
opportunity. The placement and orientation of these observatories can be optimized 
for sensitivity to Galactic source populations, as shown in this study. Future 
work will include how multimessenger analysis of Galactic sources can further improve 
prospects for probing the history and structure of the Milky Way, fundamental physics, and the physics of compact objects.

\section*{Acknowledgements}
We thank Bryan Gaensler, Juna Kollmeier, Norm Murray, Ziggy Pleunis, 
and Aaron Tohuvavohu for fruitful discussions that have greatly benefited 
this work. S.E.G. is supported by the Natural Sciences and Engineering Research 
Council of Canada (NSERC), funding reference CITA 490888-16. E.D.H. is supported by
the MathWorks, Inc. S.M.N. is grateful for financial support from the Nederlandse Organisatie voor Wetenschappelijk Onderzoek (NWO) through the Projectruimte and VIDI grants.

%%%%%%%%%
\software{python \citep{Python:a,Python:b}, 
          numpy \citep{NumPy:a,NumPy:b}, 
          matplotlib \citep{Hunter:2007},
          iwanthue \citep{iwanthue},
          ColorBrewer \citep{cbrewer},
          Color Oracle \citep{coracle}.
          }

%%% If you wish to include an acknowledgments section in your paper,
%%% separate it off from the body of the text using the \acknowledgments
%%% command.

%\appendix
%A handy "cheat sheat" that provides the necessary LaTeX to produce 17 
%different types of tables is available at %\url{http://journals.aas.org/authors/aastex/aasguide.html#table_cheat_sheet}.
%\appendix


\bibliography{gw-detectors-for-galactic-science}



\end{document}